
\input epsf
\input harvmac 
\newif\ifdraft

\noblackbox
\catcode`\@=11
\newif\iffrontpage
\def\figin{\epsfcheck\figin}\def\figins{\epsfcheck\figins}
\def\epsfcheck{\ifx\epsfbox\UnDeFiNeD
\message{(NO epsf.tex, FIGURES WILL BE IGNORED)}
\gdef\figin##1{\vskip2in}\gdef\figins##1{\hskip.5in}%
\else\message{(FIGURES WILL BE INCLUDED)}%
\gdef\figin##1{##1}\gdef\figins##1{##1}\fi}
\def\DefWarn#1{}
\def\figinsert{\goodbreak\midinsert}
\def\ifig#1#2#3{\DefWarn#1\xdef#1{fig.~\the\figno}
\writedef{#1\leftbracket fig.\noexpand~\the\figno}%
\figinsert\figin{\centerline{#3}}\medskip%
\centerline{\vbox{\baselineskip12pt
\advance\hsize by -1truein\noindent\tensl%
{{\bf Fig.~\the\figno}~#2}}
}\bigskip\endinsert\global\advance\figno by1}
\ifx\answ\bigans
\def\titleft{\titsm}
\magnification=1200\baselineskip=14pt plus 2pt minus 1pt
%
\voffset=0.5truein\hoffset=0.15truein
\hsize=6.15truein\vsize=600.truept\hsbody=\hsize\hstitle=\hsize
\else\let\lr=L
\def\titleft{\titla}
\magnification=1000\baselineskip=14pt plus 2pt minus 1pt
%
\hoffset=-0.25truein\voffset=-.0truein
\vsize=6.5truein
\hstitle=8.truein\hsbody=4.75truein
\fullhsize=10truein\hsize=\hsbody
\fi
\parskip=4pt plus 15pt minus 1pt

\font\titla=cmr10 scaled\magstep3
\font\tenmss=cmss10
\font\absmss=cmss10 scaled\magstep1

\font\twelvebf=cmbx10 scaled\magstep1

\newfam\mssfam
\font\footrm=cmr8  \font\footrms=cmr5
\font\footrmss=cmr5   \font\footi=cmmi8
\font\footis=cmmi5   \font\footiss=cmmi5
\font\footsy=cmsy8   \font\footsys=cmsy5
\font\footsyss=cmsy5   \font\footbf=cmbx8
\font\footmss=cmss8
\def\footfont{\def\rm{\fam0\footrm}
\textfont0=\footrm \scriptfont0=\footrms
\scriptscriptfont0=\footrmss
\textfont1=\footi \scriptfont1=\footis
\scriptscriptfont1=\footiss
\textfont2=\footsy \scriptfont2=\footsys
\scriptscriptfont2=\footsyss
\textfont\itfam=\footi \def\it{\fam\itfam\footi}
\textfont\mssfam=\footmss \def\mss{\fam\mssfam\footmss}
\textfont\bffam=\footbf \def\bf{\fam\bffam\footbf} \rm}
\def\tenpoint{\def\rm{\fam0\tenrm}
\textfont0=\tenrm \scriptfont0=\sevenrm
\scriptscriptfont0=\fiverm
\textfont1=\teni  \scriptfont1=\seveni
\scriptscriptfont1=\fivei
\textfont2=\tensy \scriptfont2=\sevensy
\scriptscriptfont2=\fivesy
\textfont\itfam=\tenit \def\it{\fam\itfam\tenit}
\textfont\mssfam=\tenmss \def\mss{\fam\mssfam\tenmss}
\textfont\bffam=\tenbf \def\bf{\fam\bffam\tenbf} \rm}
\ifx\answ\bigans\def\abstractfont{\tenpoint}\else
\def\abstractfont{\def\rm{\fam0\absrm}
\textfont0=\absrm \scriptfont0=\absrms
\scriptscriptfont0=\absrmss
\textfont1=\absi \scriptfont1=\absis
\scriptscriptfont1=\absiss
\textfont2=\abssy \scriptfont2=\abssys
\scriptscriptfont2=\abssyss
\textfont\itfam=\bigit \def\it{\fam\itfam\bigit}
\textfont\mssfam=\absmss \def\mss{\fam\mssfam\absmss}
\textfont\bffam=\absbf \def\bf{\fam\bffam\absbf}\rm}\fi
%
\def\f@@t{\baselineskip10pt\lineskip0pt\lineskiplimit0pt
\bgroup\aftergroup\@foot\let\next}
\setbox\strutbox=\hbox{\vrule height 8.pt depth 3.5pt width\z@}
\def\vfootnote#1{\insert\footins\bgroup
\baselineskip10pt\footfont
\interlinepenalty=\interfootnotelinepenalty
\floatingpenalty=20000
\splittopskip=\ht\strutbox \boxmaxdepth=\dp\strutbox
\leftskip=24pt \rightskip=\z@skip
\parindent=12pt \parfillskip=0pt plus 1fil
\spaceskip=\z@skip \xspaceskip=\z@skip
\Textindent{$#1$}\footstrut\futurelet\next\fo@t}
\def\Textindent#1{\noindent\llap{#1\enspace}\ignorespaces}
\def\footnote#1{\attach{#1}\vfootnote{#1}}%

\def\foot{\attach\footsymbolgen\vfootnote{\footsymbol}}
\let\footsymbol=\star
\newcount\lastf@@t           \lastf@@t=-1
\newcount\footsymbolcount    \footsymbolcount=0
\def\footsymbolgen{\relax\footsym
\global\lastf@@t=\pageno\footsymbol}
\def\footsym{\ifnum\footsymbolcount<0
\global\footsymbolcount=0\fi
{\iffrontpage \else \advance\lastf@@t by 1 \fi
\ifnum\lastf@@t<\pageno \global\footsymbolcount=0
\else \global\advance\footsymbolcount by 1 \fi }
\ifcase\footsymbolcount \fd@f\star\or
\fd@f\dagger\or \fd@f\ast\or
\fd@f\ddagger\or \fd@f\natural\or
\fd@f\diamond\or \fd@f\bullet\or
\fd@f\nabla\else \fd@f\dagger
\global\footsymbolcount=0 \fi }
\def\fd@f#1{\xdef\footsymbol{#1}}
\def\space@ver#1{\let\@sf=\empty \ifmmode #1\else \ifhmode
\edef\@sf{\spacefactor=\the\spacefactor}
\unskip${}#1$\relax\fi\fi}
\def\attach#1{\space@ver{\strut^{\mkern 2mu #1}}\@sf}
%
\newif\ifnref
\def\rrr#1#2{\relax\ifnref\nref#1{#2}\else\ref#1{#2}\fi}
\def\ldf#1#2{\begingroup\obeylines
\gdef#1{\rrr{#1}{#2}}\endgroup\unskip}
\def\nrf#1{\nreftrue{#1}\nreffalse}
\def\doubref#1#2{\refs{{#1},{#2}}}
\def\multref#1#2#3{\nrf{#1#2#3}\refs{#1{--}#3}}
\nreffalse
\def\refout{
\vskip2.truecm \immediate\closeout\rfile\writestoppt
\baselineskip=14pt\centerline{{\twelvebf
References}}\bigskip{\frenchspacing\parindent=20pt
\escapechar=` \input \jobname.refs\vfill\eject}\nonfrenchspacing}
%
\def\eqn#1{\xdef #1{(\secsym\the\meqno)}
\writedef{#1\leftbracket#1}%
\global\advance\meqno by1\eqno#1\eqlabeL#1}
\def\eqnalign#1{\xdef #1{(\secsym\the\meqno)}
\writedef{#1\leftbracket#1}%
\global\advance\meqno by1#1\eqlabeL{#1}}
%
\def\chap#1{\global\advance\secno by1\message{(\the\secno\ #1)}
\global\subsecno=0\eqnres@t\noindent{\twelvebf\the\secno\ #1}
\writetoca{{\secsym} {#1}}\par\nobreak\medskip\nobreak}
\def\eqnres@t{\xdef\secsym{\the\secno.}%
\global\meqno=1\bigbreak\bigskip}
\def\sequentialequations{\def\eqnres@t{\bigbreak}}\xdef\secsym{}
\global\newcount\subsecno \global\subsecno=0
\def\sect#1{\global\advance\subsecno
by1\message{(\secsym\the\subsecno. #1)}
\ifnum\lastpenalty>9000\else\bigbreak\fi
\noindent{\bf\secsym\the\subsecno\ #1}\writetoca{\string\quad
{\secsym\the\subsecno.} {#1}}\par\nobreak\medskip\nobreak}
\def\chapter#1{\chap{#1}}
\def\section#1{\sect{#1}}
\def\\{\ifnum\lastpenalty=-10000\relax
\else\hfil\penalty-10000\fi\ignorespaces}
\def\note#1{\leavevmode%
\edef\@@marginsf{\spacefactor=\the\spacefactor\relax}%
\ifdraft\strut\vadjust{%
\hbox to0pt{\hskip\hsize%
\ifx\answ\bigans\hskip.1in\else\hskip .1in\fi%
\vbox to0pt{\vskip-\dp
\strutbox\sevenbf\baselineskip=8pt plus 1pt minus 1pt%
\ifx\answ\bigans\hsize=.7in\else\hsize=.35in\fi%
\tolerance=5000 \hbadness=5000%
\leftskip=0pt \rightskip=0pt \everypar={}%
\raggedright\parskip=0pt \parindent=0pt%
\vskip-\ht\strutbox\noindent\strut#1\par%
\vss}\hss}}\fi\@@marginsf\kern-.01cm}
\def\titlepage{%
\frontpagetrue\nopagenumbers\abstractfont%
\hsize=\hstitle\rightline{\vbox{\baselineskip=10pt%
{\abstractfont\pubnum}}}\pageno=0}
\frontpagefalse
\def\pubnum{}
\def\pdate{\number\month/\number\yearltd}
\def\makefootline{\iffrontpage\vskip .27truein
\line{\the\footline}
\vskip -.1truein\leftline{\vbox{\baselineskip=10pt%
{\abstractfont\pdate}}}
\else\vskip.5cm\line{\hss \tenrm $-$ \folio\ $-$ \hss}\fi}
\def\title#1{\vskip .7truecm\titlestyle{\titleft #1}}
\def\titlestyle#1{\par\begingroup \interlinepenalty=9999
\leftskip=0.02\hsize plus 0.23\hsize minus 0.02\hsize
\rightskip=\leftskip \parfillskip=0pt
\hyphenpenalty=9000 \exhyphenpenalty=9000
\tolerance=9999 \pretolerance=9000
\spaceskip=0.333em \xspaceskip=0.5em
\noindent #1\par\endgroup }
\def\autskip{\ifx\answ\bigans\vskip.5truecm\else\vskip.1cm\fi}
\def\author#1{\vskip .7in \centerline{#1}}
\def\andauthor#1{\autskip
\centerline{\it and} \autskip\centerline{#1}}
\def\address#1{\ifx\answ\bigans\vskip.2truecm
\else\vskip.1cm\fi{\it \centerline{#1}}}
\def\abstract#1{
\vskip .5in\vfil\centerline
{\bf Abstract}\penalty1000
{{\smallskip\ifx\answ\bigans\leftskip 2pc \rightskip 2pc
\else\leftskip 5pc \rightskip 5pc\fi
\noindent\abstractfont \baselineskip=12pt
{#1} \smallskip}}
\penalty-1000}
\def\endpage{\tenpoint\supereject\global\hsize=\hsbody%
\frontpagefalse\footline={\hss\tenrm\folio\hss}}
\def\ack{\goodbreak\vskip2.cm\centerline{{\twelvebf
Acknowledgements}}}
\def\append#1#2{\global\meqno=1\global
\subsecno=0\xdef\secsym{\hbox{#1.}}
\bigbreak\bigskip\noindent{\twelvebf Appendix #1. #2}\message{(#1.
#2)} \writetoca{\twelvebf{Appendix {#1.} {#2}}}\nobreak}
%
\def\CERN{\address{CERN, Geneva, Switzerland}}
\def\bfone{\relax{\rm 1\kern-.35em 1}}
\def\inbar{\vrule height1.5ex width.4pt depth0pt}
\def\IC{\relax\,\hbox{$\inbar\kern-.3em{\mss C}$}}
\def\ID{\relax{\rm I\kern-.18em D}}
\def\IF{\relax{\rm I\kern-.18em F}}
\def\IH{\relax{\rm I\kern-.18em H}}
\def\II{\relax{\rm I\kern-.17em I}}
\def\IN{\relax{\rm I\kern-.18em N}}
\def\IP{\relax{\rm I\kern-.18em P}}
\def\IQ{\relax\,\hbox{$\inbar\kern-.3em{\rm Q}$}}
\def\IR{\relax{\rm I\kern-.18em R}}
\font\cmss=cmss10 \font\cmsss=cmss10 at 7pt
\def\ZZ{\relax\ifmmode\mathchoice
{\hbox{\cmss Z\kern-.4em Z}}{\hbox{\cmss Z\kern-.4em Z}}
{\lower.9pt\hbox{\cmsss Z\kern-.4em Z}}
{\lower1.2pt\hbox{\cmsss Z\kern-.4em Z}}\else{\cmss Z\kern-.4em
Z}\fi}
\def\a{\alpha} \def\b{\beta} \def\d{\delta}
 
 \def\l{\lambda}

\def\cJ{{\cal J}} 
 
 \def\cO{{\cal O}}
 
\def\cR{{\cal R}} \def\cV{{\cal V}}
\def\nup#1({Nucl.\ Phys.\ $\us {B#1}$\ (}
\def\plt#1({Phys.\ Lett.\ $\us  {#1}$\ (}
\def\cmp#1({Comm.\ Math.\ Phys.\ $\us  {#1}$\ (}
\def\prp#1({Phys.\ Rep.\ $\us  {#1}$\ (}
\def\prl#1({Phys.\ Rev.\ Lett.\ $\us  {#1}$\ (}
\def\prv#1({Phys.\ Rev.\ $\us  {#1}$\ (}
\def\mpl#1({Mod.\ Phys.\ Let.\ $\us  {A#1}$\ (}
\def\ijmp#1({Int.\ J.\ Mod.\ Phys.\ $\us{A#1}$\ (}
\def\tit#1|{{\it #1},\ }
%

%

\def\ni{\noindent}
\def\tilde{\widetilde}
\def\bar{\overline}
\def\us#1{\underline{#1}}

\def\hat{\widehat}

\def\Coeff#1#2{{#1\over #2}}
\def\Coe#1.#2.{{#1\over #2}}
\def\coeff#1#2{\relax{\textstyle {#1 \over #2}}\displaystyle}
\def\coe#1.#2.{\relax{\textstyle {#1 \over #2}}\displaystyle}
\def\half{{1 \over 2}}
\def\shalf{\relax{\textstyle {1 \over 2}}\displaystyle}

\def\notin{\hbox{{$\in$}\kern-.51em\hbox{/}}}

\def\exx#1{e^{{\displaystyle #1}}}
\def\del{\partial}

\def\nex#1{$N\!=\!#1$}

\catcode`\@=12
\def\sc{superconformal\ }

\def\LG{Lan\-dau-Ginz\-burg\ }
\def\nul#1,{{\it #1},}

\def\LG{Lan\-dau-Ginz\-burg\ }

\def\cph#1#2{\widehat{CP}_{#1}^{\lower2pt\hbox{$\scriptstyle(#2)$}}}

\def\cph#1#2{{\rm CP}_{\!#1,#2}}
\def\rx#1{\cR_x^{(#1)}}

\def\dx#1{\del_{x_{#1}}}

\def\bfone{{\bf 1}}

\def\Gm{G^-}
\def\Um{U^-}
\def\dx#1{{\del\over\del x_{#1}}}
\def\xy{(x_1,x_2)}
\def\wk#1{W_{#1}\xy}
\def\tvp{\vrule height 3.2pt depth 1pt} 
\def\thp{\vrule height 0.4pt width 0.45em}
\def\ccw#1{\hfill#1\hfill}
\setbox111=\vbox{\offinterlineskip
\cleartabs
\+ \thp&\cr
\+ \tvp\ccw{}&\tvp\cr
\+ \thp&\cr
\+ \tvp\ccw{}&\tvp\cr
\+ \thp&\cr}
\def\g{\gamma}
\def\cJ{{\cal T}}
\def\cV{{\cal W}}
\def\DT#1{{(\!D_{\!+}#1\!)}}
\def\DBT#1{{(\!D_{\!-}#1\!)}}
\def\DPM#1{{(\!D_{\!+-}#1\!)}}
\def\tp{\theta^+}
\def\tm{\theta^-}
\def\JJ{\cJ_{l,1}}
\def\lv{Liouville}
\def\I{{\cal K}}
\def\La#1{\Lambda^{\! #1}}
\def\qbrs{Q_{BRST}}
\def\Vl{{\cV_l}}
\def\Jl{{\cJ_l}}
\def\osp{O\!Sp}

%

\ldf\TOPALG{E.\ Witten, \cmp{117} (1988) 353; \cmp{118} (1988) 411;
\nup340 (1990) 281.}
\ldf\EYtop{T.\ Eguchi and S.\ Yang, \mpl4 (1990) 1693.}
\ldf\topgr{E.\ Witten, \nup340 (1990) 281; R.\ Dijkgraaf and E.\
Witten, \nup342(1990) 486; J.\ Distler, \nup342(1990) 523; E.\ and
H.\ Verlinde, \nup348 (1991) 457; R.\ Dijkgraaf and E.\ and H.\
Verlinde, \nup348 (1991) 435; For a review, see: R.\ Dijkgraaf, {\it
Intersection theory, integrable hierarchies and topological field
theory}, preprint IASSNS-HEP-91/91.}
\ldf\EHV{E.\ and H.\ Verlinde, in \topgr.}
\ldf\Witgr{E.\ Witten, \nup373 (1992) 187. }
\ldf\DVV{R.\ Dijkgraaf, E. Verlinde and H. Verlinde, \nup{352} (1991)
59.}
\ldf\Loss{A. Lossev, \nul{ Descendants constructed from matter field
and K.
Saito higher residue pairing in Landau-Ginzburg theories coupled to
topological
gravity}, preprint TPI-MINN-92-40-T.}
\ldf\BGS{B.\ Gato-Rivera and A.M.\ Semikhatov, \plt B293 (1992) 72.}
\ldf\LVW{W.\ Lerche, C.\ Vafa and N.P.\ Warner, \nup324 (1989) 427.}
\ldf\cring{D.\ Gepner, \nul{ A comment on the chiral algebras of
quotient
superconformal field theories}, preprint PUPT-1130; S.\ Hosono and
A.\
Tsuchiya, \cmp136(1991) 451.}
\ldf\Wchiral{K.\ Ito, \plt B259(1991) 73; \nup370(1992) 123; D.\
Nemeschansky and S.\ Yankielowicz, \nul{ N=2 W-algebras,
Kazama-Suzuki models and Drinfeld-Sokolov reduction}, preprint
USC-91-005A;
L.\ Romans, \nup369 (1992) 403.}
\ldf\topw{K.\ Li, \plt B251 (1990) 54, \nup346 (1990) 329, {\it
Linear $W_N$-gravity}, preprint CALT-68-1724; H.\ Lu, C.N.\ Pope and
X.\ Shen, \nup366(1991) 95; S.\ Hosono, \nul{ Algebraic definition of
topological W-gravity}, preprint UT-588; H.\ Kunitomo, Prog.\ Theor.\
Phys.\ 86 (1991) 745.}
\ldf\MS{P.\ Mansfield and B.\ Spence, \nup362(1991) 294.}
\ldf\KS{Y.\ Kazama and H.\ Suzuki, \nup321(1989) 232.}
\ldf\Keke{K.\ Li, \nup354(1991) 711; \nup354(1991)725.}
\ldf\Vafa{C.\ Vafa, \mpl6 (1991) 337.}
\ldf\noncritW{A.\ Bilal and J.\ Gervais, \nup326(1989) 222; P.\
Mansfield and B.\ Spence, \nup362(1991) 294; M.\ Bershadsky, W.\
Lerche, D.\ Nemeschansky and N.P.\ Warner, \plt B292 (1992) 35.}
\ldf\BLNWB{M.\ Bershadsky, W.\ Lerche, D.\ Nemeschansky and N.P.\
Warner,
\nup401 (1993) 304.}
\ldf\EYQ{T.\ Eguchi, H.\ Kanno, Y.\ Yamada and S.-K.\ Yang, \plt B305
(1993) 235.}
\ldf\flatco{K.\ Saito, J.\ Fac.\ Sci.\ Univ.\ Tokyo Sec.\ IA.28
(1982) 775; M.\ Noumi, Tokyo.\ J.\ Math. 7 (1984) 1; B.\ Blok and A.\
Varchenko, \tit Topological conformal field theories and the flat
coordinates| preprint IASSNS-HEP-91/5; S.\ Cecotti and C.\ Vafa,
\nup367 (1991) 359; W.\ Lerche, D.\ Smit and N.\ Warner, \nup372
(1992) 87.}
\ldf\superpot{P.\ Fendley, W.\ Lerche, S.\ Mathur and N.P.\ Warner,
\nup348 (1991) 66; W.\ Lerche and N.P.\ Warner, \nup358 (1991) 571.}
\ldf\fusionr{D.\ Gepner, \cmp 141 (1991) 381.}
\ldf\BMP{P.\ Bouwknegt, J.\ McCarthy and K.\ Pilch, Lett.\ Math.\
Phys.\ 29 (1993) 91; {\it On the BRST structure of $W_3$-gravity
coupled to c=2 matter}, preprint USC-93/14; {\it On the $W$-gravity
spectrum and its G-structure}, preprint USC-93/27.}
\ldf\wbrs{M.\ Bershadsky, W.\ Lerche, D.\ Nemeschansky and N.P.\
Warner, \plt B292 (1992) 35; E.\ Bergshoeff, A.\ Sevrin and X.\ Shen,
\plt296 (1992) 95; J. de Boer and J. Goeree, \nup405 (1993) 669.}
\ldf\wlkdv{W.\ Lerche, {\it Generalized Drinfeld-Sokolov
Hierarchies, Quantum Rings, and W-Gravity}, preprint
CERN-TH.6988/93.}
\ldf\FGLS{P.\ Fr\'e, L.\ Girardello, A.\ Lerda and P.\ Soriani,
\nup387 (1992) 333.}
\ldf\witLG{E.\ Witten, {\it On the Landau-Ginzburg description of N=2
minimal models}, preprint IASSNS-HEP-93-1.}
\ldf\Larry{L.\ Romans, \nup352(1991) 829.}
\ldf\DeNi{D.\ Nemeschanksy and N.\ Warner, {\it Refining the elliptic
genus}, preprint USC-94/002.}
\ldf\wlberk{For a review, see eg., W.\ Lerche, {\it Chiral Rings and
Integrable Systems for Models of Topological Gravity}, to appear in
the Proceedings of Strings '93, Berkeley, preprint CERN-TH.7128/93,
hep-th/9401121.}
\ldf\DiNe{J.\ Distler and P.\ Nelson, Phys.\ Rev.\ Lett.\ 66 (1991)
1955.}
\ldf\JDPN{J.\ Distler and P.\ Nelson, \cmp138 (1991) 273.}
\ldf\BLLS{A.\ Boresch, K.\ Landsteiner, W.\ Lerche and A.\ Sevrin,
{\it Topological strings from Hamiltonian reduction}, to appear.}
\ldf\supergr{ V.\ Kac, \cmp53 (1977) 31; L.\ Frappat, E.\ Ragouchy
and P.\ Sorba, \cmp157 (1993) 499.}
\ldf\hamr{J.\ de Boer and T.\ Tjin,  {\it The relation between
quantum W algebras and Lie algebras}, preprint THU-93-05; A. Sevrin
and W. Troost, \plt315 (1993) 304.}
\ldf\GS{P.\ Goddard and A.\ Schwimmer, \plt214 (1988) 447.}
\ldf\MD{M.\ Douglas, \plt238B(1990) 176.}
\ldf\LGREFS{
C.\ Vafa and N.P.\ Warner, \plt218B (1989) 51; E. Martinec, \plt
217B(1989) 431.}
%
\baselineskip=14pt plus 2pt minus 1pt
\def\pubnum{
\hbox{CERN-TH.7210/94}
\hbox{hep-th/9403183}
}
\def\pdate{}
\titlepage
\vskip 1.cm
\title{On the \LG Realization of Topological Gravities}
\vskip .3cm
\author{\ W.$\,$Lerche}
\andauthor{A.$\,$Sevrin}
\CERN
\vskip0. cm
\abstract{
We study the equivariant cohomology of a class of multi-field
topological \LG models, and find that such systems carry intrinsic
information about $W$-gravity. As a result, we can
construct the gravitational chiral ring in terms of LG polynomials.
We find, in particular, that the spectrum of such theories seems to
be richer than so far expected. We also briefly discuss
the BRST operator for non-linear topological $W$-gravity.
}
\vfil
\ni CERN-TH.7210/94\hfill\break
\ni March 1994
\endpage
\baselineskip=14pt plus 2pt minus 1pt

\chapter{Introduction}

By now it is well-known \multref\BGS\BLNWB\wlberk\ that certain
systems of conformal matter coupled to $2d$-gravity have an
equivalent description in terms of topological \LG theory \Vafa. More
precisely, the minimal models of type $(1,t)$ coupled to gravity are
closely related to the twisted \doubref\TOPALG\EYtop\ \nex2
superconformal minimal models of type $A_{k+1}$, where $t=k+2$. In
particular, the dynamics of both systems are governed
\doubref\MD\DVV\ by the same KdV-type of integrable systems. This, in
effect, allows to use \nex2 \LG methods to obtain further insight in
the matter-gravity systems, and to explicitly compute various
correlation functions.

There is, however, an important difference between the matter-gravity
system and the topological minimal model: the spectrum of of
$A_{k+1}$ is given by the chiral primary ring
$$
\cR_x\ =\ \big\{\,1,x,x^2,\dots,x^k\,\big\}\ ,
\eqn\Rx
$$
whereas for the chiral ground ring of the matter-gravity system one
has infinitely many more physical operators:
$$
\cR_{x,\g}\ =\
\cR_x\,\otimes\,\big\{\,(\g^0)^l,\,l=0,1,2,\dots\,\big\}\ .
\eqn\Rxg
$$
This is also the spectrum of an a priori different system,
namely of topological minimal matter $A_{k+1}$ coupled \Keke\ to
topological gravity \topgr. From this viewpoint, one may interpret
$(\g^0)^l$ as gravitational descendants and call $\cR_{x,\g}$ an
gravitationally extended chiral ring.

One can actually modify the cohomological definition of the
topological matter theory such that its spectrum becomes precisely
$\cR_{x,\g}$ \doubref\BLNWB\EYQ. That is, upon requiring equivariant
cohomology, which amounts to imposing the condition that the
anti-ghost $b$ annihilates the physical states \doubref\topgr\DiNe,
the gravitational descendants $(\g^0)^l$ become physical in the
modified theory. This means, however, that the structure of the
gravitational sector must, in some way, already be built in the
topological matter model.

Indeed, in the \LG formulation of topological minimal matter \LGREFS,
where $x$ in \Rx\ is viewed as the LG field (with superpotential
$W(x)=x^{k+2}$), the gravitational descendants can be very simply
constructed in terms of $x$ as well \Loss:\foot{We will denote the
(abstract) excitations of topological gravity by $(\g^0)^l$ and their
LG representatives by $\sigma_l$.}
$$
(\g^0)^lx^i\ =\ \sigma_l(x^i)\ =\ x^{i+(k+2)l}\ ,
\ \ \ \ x^{(l+1)(k+2)-1}\ \equiv\ 0\
\eqn\gx
$$
(we also exhibited here the null states of the theory). In more
technical terms, the BRST cohomology of the topological
matter-gravity system has representatives that lie entirely in the
matter sector \EYQ.

It is the gravitational ring structure \gx, depicted in Fig.1, that
is common to many algebraic features of the theory, like for example
the spectrum of KdV flows. One of our concerns is to analyze the
corresponding structure of more complicated versions of $2d$ gravity
in connection with \LG models. In essence, gravity models with
additional symmetries lead to LG models with more basic fields and
more types of gravitational excitations. In particular, we want to
understand the gravitational chiral rings that are
implicitly built in more complicated LG models. The rule, valid at
least for a large class of theories, seems to be that one gets a new
type of gravitational descendant generator for each new LG field, as
well as an independent supercurrent generator of the extended
topological symmetry algebra \doubref\BLNWB\wlkdv. What we hope for
to find, ultimately, is an intrinsic and generic relationship between
\nex2 \LG theory, variations of topological gravity and integrable
systems \wlkdv.
\ifig\figone{Chiral ring spectrum and pattern of KdV
flows in the topological minimal model of type $A_5$ coupled to
gravity. The open dot describes a null
field.}{\epsfxsize3.5in\epsfbox{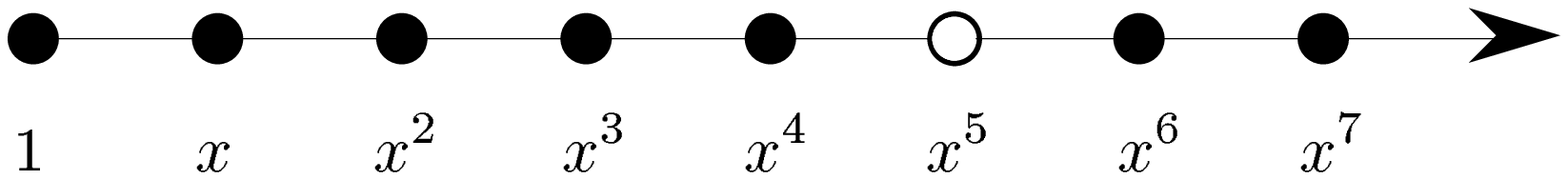}}

In the present paper, we consider as next step in this project the
question, how the extended gravitational sector can be recovered
from within a more general, multi-variable LG theory.

Specifically, we will focus on the $W_3$ extension of gravity coupled
\wbrs\ to matter of type $(1,t)$. We do this not because $W_3$
matter-gravity systems would be particularly important, but because
these systems are the simplest possible extensions of ordinary
gravity theories, the study of which ought to provide insight into
the general case. It is known \BLNWB\ that such systems have an
analogous close connection to topological $W_3$ minimal models, which
are just the (twisted) \nex2 coset models \KS\ based on $SU(3)_k\over
U(2)$ (where $t=k+3$), and which have a well-understood
\doubref\LVW\cring\ \LG description as well.

In the next section, we will study the $W_3$ symmetry properties of
such LG models. We then couple them to {\it non-linear} topological
$W_3$-gravity, which we define in terms of quantum Hamiltonian
reduction of $\osp(6|4)$ and which is different from the kinds of
{\it linear} $W_3$-gravity discussed so far in the literature \topw.
Though we did not fully succeed in explicitly writing down the
complete BRST operator, we will go sufficiently far to be able to
investigate the relevant part of the equivariant cohomology of the
matter-gravity system. We will see that the non-trivial cohomology
can indeed be represented in terms of the \LG sector alone, and we
will be able to write the observables of topological $W$-gravity in
terms of LG polynomials. As it will be discussed in the conclusions,
we find that there are actually more physical operators than it was
expected from previous work. Finally, in the appendix we will explain
our superspace conventions.

\chapter{W$_{\!{\textstyle{\bf 3}}}$ symmetry in N=2
\LG models}

Following in spirit \doubref\FGLS\witLG, we will describe the
topological $W_3$ matter sector in terms ``almost'' free LG
fields.\foot{After we had finished the computations for this section,
we received the paper \DeNi\ which has some overlap with our
results.} We find that the formulae become particularly compact if we
choose chiral superfields $X_i, \Psi_i$ $i=1,2$, as follows:\foot {We
will be considering in this paper only the left-moving part of the
theory. For the right-movers, analogous formulae hold.}
$$
\eqalign{
X_i\ &=\ x_i \,+\, \tm\psi_i^+
-\shalf\tp\tm\del x_i\cr
\Psi_i\ &=  \bar\psi{}_i^+
 + \tm\del x^*_i -\shalf\tp\tm\del\bar\psi{}_i^+\ .\cr
}\eqn\XPdef
$$
The component fields are the usual building blocks of topological LG
models \Vafa, and are nothing but a collection of ghost systems in
disguise: $x=\beta, \psi^+=b,\bar\psi{}^+=c, \del x^*=\gamma$. In
computing operator products, one can safely ignore the
superpotential, since it gives rise only to soft contributions.
Therefore, we are effectively dealing with a free theory,
$$
X_i(Z_1)\,\Psi_j(Z_2)\ =\ \delta_{ij}\,
\Coeff{\theta^-_{12}}{z_{12}}\ ,
\eqn\XPope
$$
with super stress tensor \doubref\FGLS\witLG
$$
\cJ_m(X,\Psi)\ =\
\sum_{i=1}^2\,[\, \omega_i X_{i} \DBT {\Psi_{i}}+
(\omega_i-1)\DBT {X_{i}} \Psi_{i}\,]
\
\eqn\Jdef
$$
(relevant for us is, of course, the topologically twisted version
\doubref\TOPALG\EYtop, which is considered later). Above,
$\omega_i = i/t$ are the $U(1)$ charges of the fields $X_i$. The
stress tensor gives rise to a central charge of
$$
c_m\ =\ 6\,\Big(\Coeff{t-3}t\Big)\ ,\eqn\Cntwo
$$
and this is precisely the central charge of the \nex2
coset models $\cph2k$ \KS\ based on $SU(3)_k\over U(2)$, if one sets
$$
t\ =\ k+3\ .\eqn\tdef
$$
These models are well-known to exhibit an \nex2 $W_3$ superconformal
symmetry \Wchiral, and indeed we can construct with the
above building blocks the following primary, spin-2 supercurrent:
$$
\eqalign{
\cV_m&(X_i,\Psi_i)\ \,=\ \cr &\a_m \Big\{ \coeff{\left( 2 - 9 t + 2
{t^2} \right) (t-1)}{\left( t-4 \right) t \left( 2 + t \right) }
\DBT{X_{1}} \DBT{X_{2}} \Psi_{1} \Psi_{2} - \coeff{(5 t-18)}{\left(
t-4 \right) \left( 2 + t \right) } \DBT{X_{1}} X_{2} \DBT{\Psi_{1}}
\Psi_{2} \cr
&-
 \coeff{(t-1)}{\left( t-4 \right) t} \DBT{X_{1}} X_{2} \Psi_{1}
\DBT{\Psi_{2}}- \coeff{(t-5) (2 t-1) }{2t \left( 2 + t \right) }
\DBT{X_{1}} \del\Psi_{1} - \coeff{(t-1)}{t} \DBT{X_{2}}
\del\Psi_{2} \cr
&+
 \coeff{(t-1) ( 2 {t^2}-5t-4)}{2\left( t-4 \right) t \left( 2 + t
\right) } \DBT{\del X_{1}} \Psi_{1} + \coeff{\left( 2 - 4 t +
{t^2} \right) (t-2)}{\left( t-4\ \right) t \left( 2 + t \right) }
\DBT{\del X_{2}} \Psi_{2} + \coeff{(t-5)}{2t \left( 2 + t \right) }
X_{1} \DBT{\del \Psi_{1}} \cr
&-
 \coeff{ (t-1) (3 t-8) }{\left( t-4 \right) t\ \left( 2 + t \right) }
X_{1} \DBT{X_{1}} \Psi_{1} \DBT{\Psi_{1}} + \coeff{2 - 9 t + 2
{t^2}}{\left( t-4 \right) t \left( 2 + t\ \right) } X_{1} \DBT{X_{2}}
\DBT{\Psi_{1}} \Psi_{2} \cr
&+
 \coeff{(3 t-8)}{2 \left( t-4 \right) t \left( 2 + t \right) }
(X_{1})^2 \DBT{\Psi_{1}}^2 + \coeff{1}{\left( t-4 \right) t}
X_{1} X_{2} \DBT{\Psi_{1}} \DBT{\Psi_{2}} + \coeff{1}{t} X_{2}
\DBT{\del \Psi_{2}} \cr
&-
 \coeff{(t-2)}{\left( t-4 \right) t} X_{2} \DBT{X_{2}} \Psi_{2}
\DBT{\Psi_{2}} + \coeff{1}{\left( t-4 \right) t} (X_{2})^2
\DBT{\Psi_{2}}^2  \cr
&-
 \coeff{ \left( 1 + t \right) ( 2 {t^2}-5t-4) }{2t\left( t-4 \right)
\left( 2 + t \right) }\, \del X_{1} \DBT{\Psi_{1}} - \coeff{\left(
2 - 4 t + {t^2} \right) }{\left( t-4 \right) t}\, \del X_{2}
\DBT{\Psi_{2}}
\Big\}\ + \, A\, \tilde\cV_m\ ,}
\eqn\Vdef
$$
where
$$
\tilde\cV_m\ =\
X_{2}\DBT{\Psi_{1}}^2+(1-t)\DBT{X_{2}}\Psi_{1}\DBT{\Psi_{1}} \ ,
\eqn\cVdef
$$
and where $A$ is a free parameter. Together with $\cJ_m$, $\cV_m$
generates the \nex2 $W_3$ \sc\ algebra (A.4). Taking
${\a_m}\equiv\sqrt{\coeff{\left( t-4 \right) \left(t+2 \right) }
{\left( 2 t-3 \right) \left( 5 t-18 \right) }}$ gives the
normalization: $\cV_m\cdot\cV_m\sim \coeff16
c_m\coeff1{(z-w)^4}+\dots$.

The full algebra that $\cJ_m$ and $\cV_m$ generate is actually
somewhat larger, due to the presence of the free parameter $A$ in
\Vdef. Indeed $\cJ_m, \tilde\cV_m$ form a separate linear \nex2 $W_3$
algebra with $\tilde\cV_m\cdot\tilde\cV_m\sim 0$, so that the full
algebra is a semi-direct product of a linear with a non-linear \nex2
$W_3$ algebra. However, the chiral algebra currents must commute with
the screening operators of the theory, a quantum version of the
requirement that the currents be conserved, and this is what fixes
$A$. In the present \nex2 \LG theories describing the models $\cph
{2}k$, the relevant screeners \doubref\FGLS\witLG\ are given in terms
of the superpotentials:
$
Q_W = \Coeff1{2\pi i}\!\oint\!dz\,d\theta^-\,W_k(X_1,X_2)
$. 
Requiring
$$
\coeff1{2\pi i}\!\oint\!dz\,d\theta^-\,W_k(X_1,X_2)(Z)\cdot \cJ_m\ =\
0\
\eqn\screenJ
$$
just yields the usual condition for $W_k$ to be quasi-homogenous:
$W_k=\sum\omega_ix_i\del_iW_k$. But requiring in addition
$$
\coeff1{2\pi i}\oint\!dz\,d\theta^-\,W_k(X_1,X_2)(Z)\cdot \cV_m\ =\
0\
\eqn\screenV
$$
gives
$$
A\ =\ \Coeff{\a_m}{t} \Coeff{(18-5t)}{(t-4)(2+t)}\ ,\eqn\Adef
$$
and produces a large list of additional differential equations for
the superpotentials $W_k$. A typical equation is, for example
$$
\eqalign{
\Big\{ &\left( t - 4 \right) \left( 2 + t \right) \dx2 + \left( 3 t -
8 \right) x_2 {\del^2\over\del {x_2}^2} + \cr &\left( 4 t - 13
\right) x_1 {\del^2\over\del {x_1}\del x_2} + \left( 5 t - 18 \right)
{\del^2\over\del {x_1}^2}\Big\}\,\wk k\ =\ 0\ .
}\eqn\sampleq
$$
The LG superpotentials of the coset models $\cph {2}k$ are well-known
\refs{\LVW{,}\superpot{,}\fusionr}, and can be compactly
characterized by the following generating function \fusionr:
$$
-\log\Big[\sum_{i=1}^{2}(-\l)^ix_i\,\Big]\ =\
\sum_{k=-2}^{\infty}\l^{k+3}\,\wk k\
\eqn\genallW
$$
Using identities like \wlkdv
$$
\eqalign{
\wk k\ \ &=\ \Coeff1{t}(2\dx1+x_1\dx2)\wk{k+1}\cr
\dx1\wk k\ &=\ (x_1\dx1+x_2\dx2)\wk{k-1}\cr
\dx2\wk k\ &=\ -\dx1\wk{k-1}\ ,\cr
}\eqn\recVR
$$
we find that all equations are solved precisely by the
superpotentials $W_k$. This proves that the models $\cph {2}k$ are
indeed associated with the ``correct'' points in the LG moduli
spaces that are determined in terms of the cohomology of
grassmannians \refs{\LVW{,}\superpot{,}\cring{,}\fusionr}.

The explicit form of the \nex2 $W_3$ symmetry currents allows us to
choose a convenient, distinguished basis for the LG polynomials,
namely the basis of eigenstates of $V_0\equiv\int\! z\,\cV_m$. We
find
that it is simply given by
$$
\Phi^{l,m}\xy\ =\ {x_2}^m\Big(\dx1\wk{l-2}\Big)\ .
\eqn\basis
$$
These specific polynomials \wlkdv\ can also be associated with the
weights of the ${\lower2pt\copy111}\,$-representation of
$SU(k\!+\!2)$. The chiral primary ring of the matter model $\cph
{2}k$ \doubref\LVW\cring\ in this basis is then
$$
\rx k\ =\ \ \Big\{\,\Phi^{l,m}\xy\ ,\ \ m+l\leq k\,\Big\}\ .
\eqn\xring
$$
We will show later how the $\Phi^{l,m}$ with $m+l>k$ can be used to
describe the $W$-gravity descendants as well.

\ni The $U(1)$ quantum numbers of the ring elements are
$$
J_0(\Phi^{l,m})=\Coeff1t(l+2m)\ ,
\eqn\charge
$$
and it is straightforward to compute the $V_0$ quantum numbers
for any given $\Phi^{l,m}$. In particular, we have
$$
\eqalign{
V_0(\Phi^{1,0}\equiv x_1)\ &=\  {\a_m}\,\Coeff{5-t}{2t(t+2)}\cr
V_0(\Phi^{0,1}\equiv x_2)\ &= -{\a_m}\,\Coeff1t \ .
}\eqn\vQN
$$
This indeed coincides with the $V_0$ quantum numbers one gets
from the explicit expressions \BLNWB\ of the ground ring generators
$x_1,x_2$ of $(1,t)$-type $W_3$-matter-gravity systems, by using
the current $V$ given in eq.\ (2.31) of ref.\ \BLNWB.
This is another confirmation that the topologically twisted
models $\cph {2}k$ are, essentially, equivalent to these systems.

We now turn to the topologically twisted version of the matter
theory. Starting from the \nex2 structure, one obtains the chiral
algebra of the topological matter system by twisting the \nex2
currents. In superspace, these currents are simply given by
$D_-\cJ_m|_{\theta^-=0}$ and $D_-\cV_m|_{\theta^-=0}$. Each of them
describes a BRST doublet with the BRST charge given by
$$
Q_S =\ \Coeff1{2\pi i}\!\oint\!dz\,d\theta^-\, \cJ_m.
$$
{}From Eq.\ (A.4), one deduces that that the algebra generated by
$D_-\cJ_m|_{\theta^-=0}$ and $D_-\cV_m|_{\theta^-=0}$ indeed closes,
but in a non-linear way. In fact, a detailed inspection reveals that
the structure functions do depend on the full \nex2 currents and not
only on the topological currents. In the next section we will
describe the coupling the matter system to topological gravity.

\chapter{Non-linear topological W$_{\!{\textstyle{\bf 3}}}$-gravity}

In \topgr, topological gravity was constructed from gauge fixing a
theory with trivial action of the $d=2$ Poincare group, a contraction
of $Sl(2)$. The resulting field content was given by \lv\ fields
$\pi$ and $\phi$, fermions $\psi$ and $\chi$ and ghosts $b$, $c$,
$\beta$ and $\gamma$. Several generalizations to topological
$W_3$-gravity were made in \topw\ by starting from a gauge theory of
some contraction of $Sl(3)$. These approaches were however
unsatisfying in the sense that the resulting twisted \nex2 $W_3$
algebra had a linear nature: the OPE of $D_-\cV$ with itself
vanished. On the other hand, from our experience with non-critical
$W$-strings \wbrs\ we rather expect the Liouville sector to
have the same type of chiral algebra as the matter sector. Though it
might very well be that the full non-linear \nex2 $W_3$ algebra
arises from the approach of \topw\ by making a different gauge
choice, we do not pursue this approach here.

Instead, motivated by a hidden, doubly twisted \nex4\ superconformal
symmetry in topological gravity \BLLS, we will rather {\it define}
non-linear topological $W$-gravity via quantum Hamiltonian reduction
of a WZW model; by general results of Hamiltonian reduction \hamr,
such a theory is consistent by construction, and the existence of a
quantum BRST operator is automatic. As will be explained below, this
approach indeed implies the existence of a non-linear \nex2
$W$-algebra in the \lv\ sector.\foot{The linear $W$-currents of
\topw\ are like the nilpotent current $\tilde\cV_m$ in \cVdef, to be
seen in contrast with the non-linear current $\cV_m$ in \Vdef.}
Therefore, we start by directly constructing the \nex2 $W_3$ algebra
out of the field content of refs.\ \topw. This essentially amounts to
doubling the fields appearing in ordinary topological gravity. Again
\nex2 superspace provides the most economical way to do so, and we
thus introduce free chiral superfields $\Pi_i$ and $\Psi_i$,
$i\,=\,1,\,2$,
$$
\eqalign{
\Pi_i\ &=\ \pi_i + \tm\del\chi_i -\shalf\tp\tm\del \pi_i\cr \Phi_i\
&=\ \psi{}_i + \tm\del\phi_i -\shalf\tp\tm\del\psi_i\ ,\cr
}\eqn\PPdef
$$
with OPE's
$$
\Pi_i(Z_1)\,\Phi_j(Z_2)\ =\
\Coeff1{z_{12}}\,\delta_{ij}\,\theta^-_{12}\ .
\eqn\PPope
$$
The super stress tensor of the ordinary topological \lv\ sector
(before twisting) is
$$
\JJ\ =\ -\Phi_{1}\,\DBT{\Pi_{1}} + q\,\DBT{\Phi_{1}}
+ q\,\del\Pi_{1}\ ,
\eqn\ordinLV
$$
which has central charge $c_{l,1}=3(1 +2 q^2)$. It turns out that
the complete \nex2 $W_3$ algebra can be realized in terms of $\JJ$
alone, together with the extra $W_3$ \lv\ fields $\Pi_2, \Phi_2$
(completely analogous to the free field realization of the ordinary
$W_3$-algebra \Larry). In particular, the total \lv\ super stress
tensor
is
$$
\cJ_l\ =\ \JJ - \Phi_{2}\,\DBT{\Pi_{2}} + q\,\DBT{\Phi_{2}} +
2\,q\,\del\Pi_{2}\ ,
\eqn\totalLV
$$
whose central charge depends on the background charge parameter $q$
as follows:
$$
c_l\ =\ 6\,(1 + 3 q^2)\ .\eqn\cLV
$$
In addition, we have the following primary, spin-2 supercurrent in
the
\lv\ sector:
$$
\eqalign{
\cV_l(&\Pi,\Phi)\ =\ \cr &\a_l \Big\{
\shalf q^2\JJ' + \coeff{2\left( 1 - 2{q^2} \right) \left( 1 + 3{q^2}
\right) }{10 + 36{q^2}}\DPM{\JJ} - \coeff{(8{q^2}-3)}{10 +
36{q^2}}(\JJ )^2
\cr&+
  \coeff{2q\left( 1 + 5{q^2} \right) }{5 + 18{q^2}}
   \JJ\DBT{\Phi_{2}} - \coeff{2\left( 1 + 5{q^2} \right) }{5 +
    18{q^2}}\JJ\DBT{\Pi_{2}}\Phi_{2} +
  \coeff{q(2{q^2}-1)}{5 + 18{q^2}}\JJ\del\Pi_{2}
\cr&+
  \coeff{q\left( 3 + 8{q^2} \right) }{5 + 18{q^2}}
   \DBT{\Pi_{2}}\Phi_{2}\DBT{\Phi_{2}} +
  \coeff{1 + 6{q^2} + 12{q^4}}{5 + 18{q^2}}
   \DBT{\Pi_{2}}\del\Phi_{2} -
  \coeff{{q^2}\left( 3 + 8{q^2} \right)
    }{2\left( 5 + 18{q^2} \right) }\DBT{\Phi_{2}}\DBT{\Phi_{2}}
\cr&+
  \coeff{\left( 1 + 3{q^2} \right) (2{q^2}-1)}{5 + 18{q^2}}
   \DBT{\del\Pi_{2}}\Phi_{2} - q\DT{\JJ}\DBT{\Pi_{2}} +
  \coeff{\left( 1 - 2{q^2} \right) \left( 1 + 2{q^2} \right) }{5
    + 18{q^2}}\del\Pi_{2}\DBT{\Phi_{2}}
\cr&+
  \coeff{q\left( 1 - 2{q^2} \right) }{5 + 18{q^2}}
   \del\Pi_{2}\DBT{\Pi_{2}}\Phi_{2} +
  \coeff{{q^2}(2{q^2}-1)}{5 + 18{q^2}}
   \del\Pi_{2}\del\Pi_{2} +
   \coeff{q\left( 1 - 2{q^2} \right) \left( 1 + 3{q^2} \right)
    }{5 + 18{q^2}}\del^{2}\Pi_{2}
\cr&-
\coeff{ q\left( 1 + 6{q^2} + 12{q^4} \right)
    }{2\left( 5 + 18{q^2} \right) }\DBT{\del\Phi_{2}}\Big\}\ .
}\eqn\Vldef
$$
which generates, together with $\cJ_l$, the non-linear \nex2 $W_3$
algebra (A.4). We stress once more that this is unlike the spin-2
\lv\ supercurrents discussed in the literature \topw, which generate
linear \nex2 $W_3$ algebras and square to zero. Taking
$\a_l=\sqrt{{(5 + 18\,{q^2})\over {\left( 1 - 2\,{q^2} \right)
\,\left( 2 + 3\,{q^2} \right) \, \left( 1 + 4\,{q^2} \right) }}}$
gives the normalization: $\cV_l\cdot\cV_l\sim \coeff16
c_l\coeff1{(z-w)^4}+\dots$. The topological currents are given by
$D_-\cJ_l |_{\theta^-=0}$ and $D_-\cV_l |_{\theta^-=0}$ and generate
precisely the same algebra as $D_-\cJ_m |_{\theta^-=0}$ and $D_-\cV_m
|_{\theta^-=0}$.

\ni The ghost system is described by free anti-chiral superfields
($i=1,2$)
$$
\eqalign{
 B_i\ &=\ \b_i + \tp b_i -\shalf\tp\tm\del \b_i\cr C_i\
&=\ c_i + \tp\g_i -\shalf\tp\tm\del c_i\ .\cr
}\eqn\BCdef
$$
with
$$
B_i(Z_1)\,C_j(Z_2)\ =\ \Coeff1{z_{12}}\,\delta_{ij}\,\theta^+_{12}\
.\eqn\BCope
$$
Above, $(b_i,c_i)$ are the usual fermionic ghosts with spins
$(i+1,-i)$, and $(\b_i,\g_i)$ are their superpartners.
The ghost number current looks
$$
\cJ_{g\#}\ =\
\sum_{i=1}^2\big\{D_{\!+}(C_iB_i)+(D_{\!+}C_i)B_i\big\}\ ,
\eqn\totalghost
$$
whereas the ghost super stress tensor is given by
$$
\cJ_{gh}\ =\ -\DT{B_{1}}\,C_{1} -2\,B_{1}\,\DT{C_{1}} -
2\,\DT{B_{2}}\,C_{2} - 3\,B_{2}\,\DT{C_{2}} \ ,
\eqn\Jghdef
$$
with central charge $c_{gh}=-3(3+5)=-24$.

In \cLV, the value $q=1$ gives $c_l=24$, which cancels the central
charge of the topological ghost system, and is the $W_3$ analog of
the ``critical'' central charge $c_l=9$ of ordinary topological
gravity. However, there is no need to cancel the central charge of
the ghosts, because in topological gravities the BRST operator
squares to zero independently of that; we will thus assume $q=1$
henceforth, even when we will couple in topological matter.

\ni In pure topological $W_3$-gravity, the BRST operator has the form
$$
\qbrs\ =\ Q_S + Q_V\ ,
\eqn\qbrsdef
$$
where
$$
Q_S\ =\ \coeff1{2\pi i}\!\oint\!dz\,d\theta^+\,\big[\,\cJ_l
+\cJ_{gh}\big]
\eqn\Qsdef
$$
is the total supersymmetry charge, and
$$
Q_V\ =\ \coeff1{2\pi i}\!\oint\!dz\,d\theta^+\,
\big[\,C_1\,D_{\!-}(\cJ_l + \shalf \cJ_{gh})\, + C_2\,(D_{\!-}\cV_l +
\shalf \hat\cV{}_{gh})\,\big]\ .
\eqn\Qvdef
$$
The only unknown in \Qvdef\ is the twisted spin-2
ghost current $\hat\cV{}_{gh}=D_{\!-}\cV{}_{gh}$. It is given to
leading order by
$$
\eqalign{
\hat\cV{}_{gh}\ &=\   3 {B_2} \del{C_1}  +\del\!{B_2} {C_1}
\cr&-
\coeff{5 i {\sqrt{2}}}{{\sqrt{69}}} {B_1} \del^{2}\!{C_1} +
\coeff{100 \left( 31 + 4 {\sqrt{6}} \right) }{5175} {B_1}
\del^{3}\!{C_2} + 4i \coeff{\sqrt2-3\sqrt3}{\sqrt{69}}\del\!{B_2}
\del{C_2} + \coeff{10 (5 {\sqrt{6}}-67)}{1725} \del^{2}\!{B_1}
\del{C_2}
\cr&+
 \coeff{4 i}{{\sqrt{23}}} \Jl {B_2} \del{C_2} - \coeff{10 (5
{\sqrt{6}}-6)}{575} \Jl \del\!{B_1} \del{C_2} + \coeff{44 i}{5
{\sqrt{23}}} \Vl {B_1} \del{C_2} + \coeff{2 (25 {\sqrt{6}}-94)}{1725}
\Jl' {B_1} \del{C_2}
\cr&+
 \coeff{2 i {\sqrt{2}}}{{\sqrt{69}}} \DBT{\Jl} {B_2} \DT{{C_2}} +
\coeff{2 i (5 {\sqrt{6}}-36)}{15 {\sqrt{23}}} \DBT{\Vl} {B_1}
\DT{{C_2}} + \coeff{1004}{1725} \DPM{\Jl} {B_1} \del{C_2}
\cr&-
\coeff{22}{575} {\Jl}^2 {B_1} \del{C_2}  + \ \dots
}\eqn\Wgh
$$
Unfortunately, for technical reasons we have not been able, despite
much effort, to compute all terms of $\hat\cV{}_{gh}$. However, we
were careful to ensure that our results, and the conclusions we draw
from them, do not depend on the extra terms in \Wgh. We also found
that already with the above, incomplete form of \Wgh, many terms
cancel in $(\qbrs)^2$ in a highly non-trivial way (actually, all
terms that are independent of the anti-ghosts $B$ and in particular,
all non-linear terms that arise from $(C_2D_{\!-}\cV_l)^2$).

The only feature that is important in the present context is that the
full BRST operator really exists at all (and satisfies the identity
(4.1) below).

In order to argue so, note that in ordinary topological gravity the
presence of two BRST charges $Q_V$, $Q_S$ (and thus of two
``BRST-ancestors'' of the energy-momentum tensor, namely the
anti-ghost $b_1$ and the fermionic current $G_-$), indicates a
hidden, doubly twisted \nex4 superconformal structure. An explicit
realization of the theory can be obtained by the Hamiltonian
reduction of $\osp(4|2)$ determined by its maximal regular subalgebra
$Sl(2|1)$. The gradation used in the reduction is tuned precisely
such that the resulting \nex4 algebra is doubly twisted. Topological
gravity from this Hamiltonian reduction point of view will be
discussed in detail in ref.\BLLS.

The obvious question in the present context is whether
the \nex4 superconformal structure of topological gravity extends to
topological $W$-gravity. It is clear that a necessary requirement for
this is the existence of an \nex4 $W_n$ algebra that contains the
\nex2 $W_n$-algebra. In order to track down such an algebra,
Hamiltonian reduction is again the method of choice. The \nex2 $W_n$
algebra is obtained from a reduction of $Sl(n|n\!-\!1)$ determined by
the principal embedding of $Sl(2|1)$ \Wchiral. Analyzing possible
embeddings of $Sl(n|n\!-\!1)$ in supergroups \supergr, one arrives at
the conclusion that only one natural candidate for an \nex4 $W_n$
algebra exists: $\osp(2n|2(n\!-\!1))$, which has $Sl(n|n\!-\!1)$ as a
maximal regular subalgebra; see Fig.2. The embedding $Sl(2|1)
\hookrightarrow \osp(2n|2(n\!-\!1))$ is the principal embedding in
the maximal regular $Sl(n|n\!-\!1)$ subalgebra. Because of this
particular choice of the embedding of $Sl(2|1)$, we have already the
guarantee that the superconformal algebra contains the \nex2 $W_n$
algebra as a subalgebra. \ifig\figdynk{The Dynkin diagrams of
$Sl(n|n\!-\!1)$ and $\osp(2n|2(n\!-\!1))$. The embedding of
$Sl(n|n\!-\!1)$ is obtained by removing the dot denoted by $\times$
in the diagram of $\osp(2n|2(n\!-\!1))$.}
{\epsfxsize3.0in\epsfbox{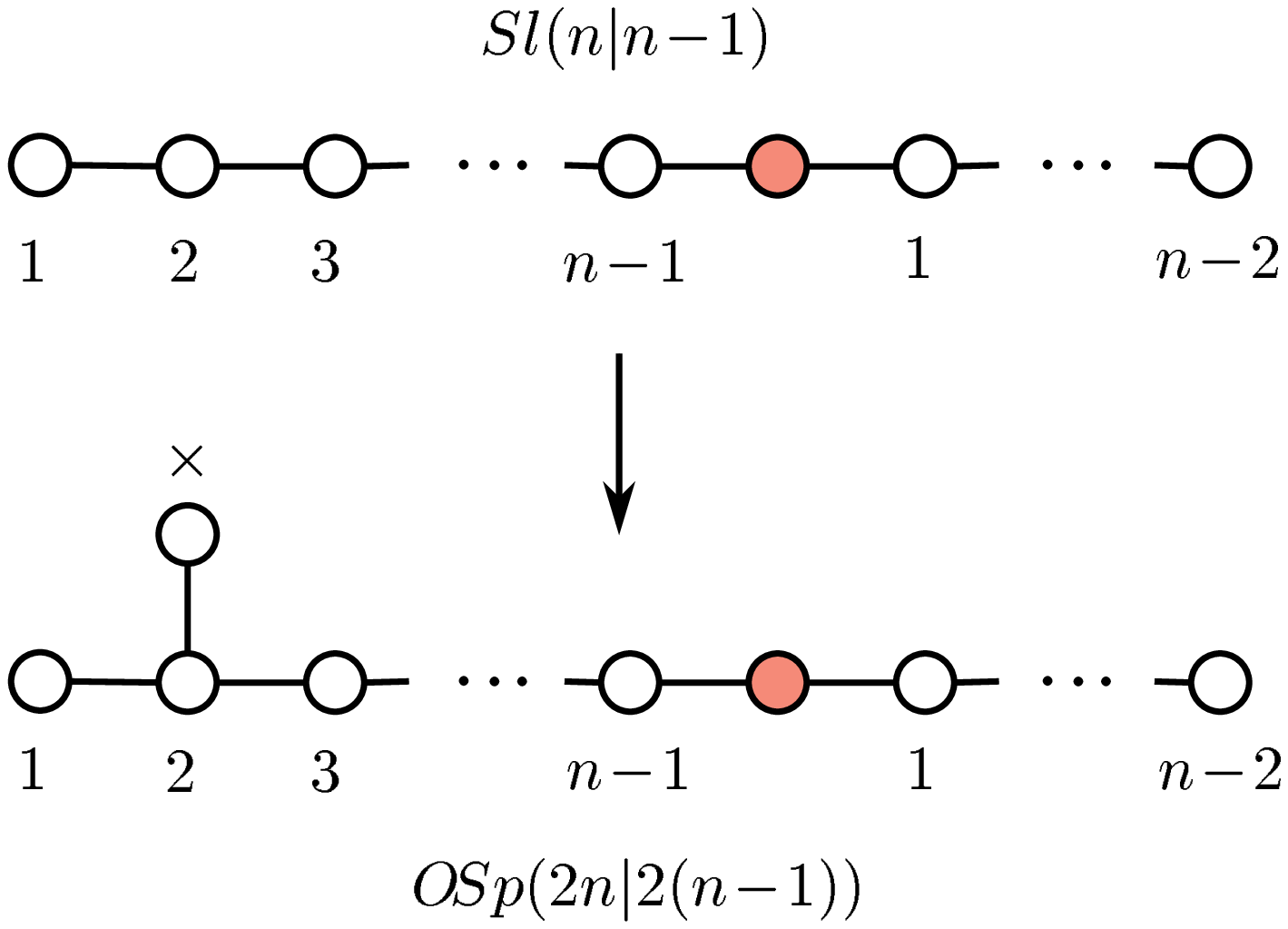}}

Using the general framework set up in \hamr, one finds that the
resulting current algebra consists of unconstrained \nex2
superfields\foot{In order to complete the ``short'' \nex2 multiplets,
some free fermions and a free scalar field have to be added. This is
the reverse of the mechanism discussed in \GS.} of dimension 1, 2,
..., $n\!-\!1$, which generate an \nex2 $W_n$ algebra. Besides these,
one has a dimension 0 superfield (which in analogy with the ordinary
\nex4 algebra only appears through its $D_+$ or $D_-$ derivative) and
couples of fields of dimension $(2k+1)/2$ where either $k=0,\ 2,
...,\ (n-2)$ for $n$ even or $k=1,\ 3, ...,\ (n-2)$ for $n$ odd. The
algebra has indeed four supersymmetry currents of dimension 3/2.
However, in order to call this an \nex4 algebra, we also need that
the supercharges can be viewed as ``square roots'' of the
translations. A sufficient condition for this is that the $Sl(2)$
subalgebra of $\osp(2n|2(n\!-\!1))$, which yields the energy-momentum
tensor in the reduction, together with the two angular momentum 1/2
multiplets (which give rise to the two other dimension 3/2
supercurrents), generate an $Sl(2|1)$ subalgebra. We verified that
this is indeed the case for $n=2$ and 3 and expect it to be valid for
arbitrary $n$.

Thus, all ingredients for performing a double twist and thus
obtaining topological $W_n$-gravity are indeed present. Though we
leave the detailed investigation of these matters for future work
\BLLS, the arguments just given almost prove the existence of $Q_V$
in topological $W$-gravity (at the quantum level). In view of the
great computational difficulties we encountered in directly
constructing $Q_V$, it might very well be that the above discussed
method provides the only economic way to obtain $Q_V$ explicitly.

In order to couple in additional matter, one may follow the
philosophy of \wbrs. There, at least for $W_3$ gravity, it was argued
that coupling matter to a non-linear algebra is possible provided the
combined matter-gravity sector has a closed (possibly soft) gauge
algebra at the classical level. The structure of the classical \nex2
$W_3$ algebra is schematically given by
$$
\eqalign{
\cJ\,\cJ\,=&\, \big[ \cJ\big]\cr
\cJ\,\cV\,=&\, \big[ \cV\big]\cr
\cV\,\cV\,=&\, \big[ \cJ\cJ\cJ\big]+ \big[ \cJ\cV\big]\ .\cr
}\eqn\wclass
$$
If we now introduce the total currents $\cJ_{tot}=\cJ_l+\cJ_m$ and
$\cV_{tot}=\cV_l-\cV_m$, we find that the classical algebra indeed
closes. The only non-trivial OPE is the one of $\cV_{tot}$ with
itself, which has the structure:
$$
\eqalign{
\cV_{tot}\,\cV_{tot}\,=\, &\big[ (\cJ_m\cJ_m-
\cJ_m\cJ_l+\cJ_l\cJ_l)\, \cJ_{tot}\big]+\cr & \shalf \big[
(\cJ_l-\cJ_m )\, \cV_{tot}\big] + \shalf \big[ (\cV_l+\cV_m )\,
\cJ_{tot}\big]\ .
}\eqn\opVV
$$
We thus expect the BRST charge to have the form:
$$
Q_V\ =\ \coeff1{2\pi i}\!\oint\!dz\,d\theta^+\,
\big[\,C_1\,D_{\!-}(\cJ_l +\cJ_m + \shalf \cJ_{gh})\, +
C_2\,(D_{\!-}\cV_l - D_{\!-}\cV_m +\shalf \hat\cV{}_{gh})\,\big]\ ,
\eqn\Qvdeflm
$$
where $\hat\cV{}_{gh}$ is very similar to eq.\ \Wgh, except
for some multiplicative renormalizations and the fact that the
Liouville-dependent parts will be replaced by the structure functions
(cf., \opVV) of the combined matter-gravity gauge algebra.
Note, though, that we do not need in the following the exact
form of \Qvdeflm. The point is that the \lv\ sector is necessary only
to obtain a covariant formulation \EHV, and can in principle be
decoupled. Indeed, it does not play any role in our subsequent
discussion of equivariant cohomology. Therefore, for our purposes, we
can effectively take as BRST operator the expression \Qvdef, with
$\cJ_l$ and $\cV_l$ replaced by $\cJ_m$ and $-\cV_m$ of eqs.\ \Jdef,
\Vdef, respectively.

\chapter{Equivariant cohomology in the topological LG system}

\ni In analogy to ref.\ \EYQ, we introduce the following
homotopy operator,
$$
S\ =\ \exx{\coeff1{2\pi i}\!\oint\!
 dz\,(c_1 \tilde G^-\! + c_2 \tilde U^-\!)}\  ,
\eqn\homotopy
$$
where
$$
\eqalign{
\tilde \Gm\ &=\ \Gm_m + \shalf \Gm_{gh}  \cr
\tilde \Um\ &= -\Um_m + \shalf \Um_{gh}\ ,\cr
}\eqn\tildef
$$
with $\Um=D_{\!-}\cV|_{\tp,\tm=0}$. It is straightforward, but
tedious to check that $S$ has the crucial property\foot{We checked
this identity up to the order for which the approximation of the BRST
operator \Qvdef, \Wgh\ is reliable. It seems to trace back to an
automorphism of $\osp(6|4)$, as will be discussed elsewhere \BLLS.}
$$
S\,(Q_S + Q_V)\,S^{-1}\ =\ Q_S\ ,
\eqn\intrig
$$
which means that the cohomology of the full BRST operator is
isomorphic to the cohomology of the supersymmetry charge,
$Q_S\equiv{1\over2\pi i}\oint (G^+_m+G^+_{gh})$. This, ultimately,
will imply that the cohomology of the topological matter-gravity
system has representatives purely in the matter sector.

Specifically, consider some polynomial in the LG fields
that is proportional to the vanishing relations,
$$
\Phi(x_1,x_2)\ =\
\sum_{i=1}^2\, P^\Phi_i(x_1,x_2)\,{\del\over\del x_i}W(x_1,x_2)\ ,
\eqn\phinull
$$
which is a null field in the topological matter model. Using
$\d_{s}\bar\psi{}^+_i = \shalf\del_{x_i}W$ for the BRST variation in
the topological matter model \refs{\Vafa{,}\FGLS{,}\EYQ}, one can
write
$$
\eqalign{
\Phi(x_1,x_2)\ &=\ \shalf\big\{\,Q_{s},\,\sum_{i=1}^2 \bar\psi{}^+_i
\,P^\Phi_i(x_1,x_2)\,\big\}\cr &= :
\shalf\big\{\,Q_{s},\,\La\Phi\,\big\}\ .
}\eqn\phiQ
$$
Whether $\Phi$ is a null operator or not after coupling to
topological gravity, depends thus on whether $\La\Phi$ is a physical
operator or not. The point is that in equivariant cohomology, the
physical states are required to be killed by the zero modes of the
anti-ghosts, and this means for $W_3$ gravity:
$(b_1)_{\,0}^-|\Phi\rangle \equiv (b_1-\bar b_1)_{\,0}|\Phi\rangle =
0$, $(b_2)_{\,0}^-|\Phi\rangle \equiv (b_2-\bar
b_2)_{\,0}|\Phi\rangle = 0$. These conditions are important for
defining correlation functions of topological gravity unambiguously
\doubref\topw\JDPN. Like for ordinary topological gravity \EYQ, one
can employ the similarity transformation $S$ to rotate the states
into the "matter" picture, $|\Phi\rangle_m = S|\Phi\rangle$.
Under this transformation, one has:
$$
\eqalign{
S\,(b_1)_{\,0}\,S^{-1}\ &=\ (b_1 + \Gm_{tot})_{\,0}\ \cr
S\,(b_2)_{\,0}\,S^{-1}\ &=\ (b_2 + \Um_{tot})_{\,0}\ ,\cr
}\eqn\shift
$$
so that in the matter picture the physical state conditions
are
$$
\eqalign{
Q_S\,|\Phi\rangle_m\ &\cong\ 0 \cr
(b_1 + \Gm_{tot})_{\,0}^-|\Phi\rangle_m\ &\cong\ 0\cr
(b_2 + \Um_{tot})_{\,0}^-|\Phi\rangle_m\ &\cong\ 0\ ,\cr
}\eqn\modequiv
$$
where ``$\, \cong\, $'' means equality modulo null fields.
This means that the naive matter null field $\Phi(x_1,x_2)$ in
\phiQ\ remains null after coupling to topological gravity
precisely if
$$
\eqalign{
\big\{\,(b_1 + \Gm_{tot})_{\,0}^-,\,\La\Phi\,\big\}\ &\cong\
0\cr \big\{\,(b_2 + \Um_{tot})_{\,0}^-,\,\La\Phi\,\big\}\
&\cong\ 0\ .
}\eqn\Lamcondx
$$
Here, only the matter parts in $\Gm_{tot}, \Um_{tot}$ can contribute,
and the matter contributions from $\Um_{gh}$ can be seen not to yield
any additional information. Therefore, the computation boils down to
determining
$$
\eqalign{
\I_1(\Phi)\ &=\ \big\{\, \Gm_{m,0},\,\La\Phi\,\big\}\cr \I_2(\Phi)\
&=\ \big\{\, \Um_{m,0},\,\La\Phi\,\big\}\ .\cr
}\eqn\Idef
$$
This involves only quantities that pertain to the topological matter
theory. Using the component forms
$$
\eqalign{
\Gm_m\ &=\ \sum \psi^+_i\del x_i^*\cr
\Um_m\ &=\
{\a_m}\Big\{2 \psi^+_{1} \psi^+_{2} {\bar\psi}{}^+_{2} \del x^*_{1} +
\coeff{18-5t} {\left( t-4 \right) \left( t+2 \right) }\psi^+_{2}
{{\del x^*_{1}}^2} + \coeff{3 t-8}{\left( t-4 \right) \left( t+2
\right) } \psi^+_{1} x_{1} {{\del x^*_{1}}^2} \cr - &\coeff{2 (t-5)
(t-1)}{\left( t-4 \right) \left( t+2 \right)\ } \psi^+_{1} \psi^+_{2}
{\bar\psi}{}^+_{1} \del x^*_{2} - \coeff{2 (t-5)}{\left( t-4 \right)
\left( t+2 \right) } \psi^+_{2} x_{1} \del x^*_{1} \del x^*_{2} +
\coeff{2 (3 t-8)}{\left( t-4 \right) \left( t+2 \right) } \psi^+_{1}
x_{2} \del x^*_{1} \del x^*_{2} \cr + &\coeff{1}{t-4} \psi^+_{2}
x_{2} {\del x^*_{2}}^2\! - \! \coeff{ 2 {t^2}-5t-4}{\left( t-4
\right) \left( t+2\ \right) } \del x^*_{1} \del \psi^+_{1}\!-\!
\coeff{2 \left( 2 - 4 t + {t^2} \right) }{\left( t-4 \right) \ \left(
t+2 \right) } \del x^*_{2} \del \psi^+_{2}\! +\! \coeff{t-5}{t+2}
\psi^+_{1} \del^2\! x^*_{1}\! +\! \psi^+_{2} \del^2\! x^*_{2} \Big\},
}\eqn\compform
$$
one can then rewrite the conditions for $\Phi(x_1,x_2)$ to
be null in the following way:\goodbreak
$$
\eqalign{
\I_1(\Phi)&=\coeff1{2\pi i}\!\oint\!
dz\,z\,\Gm_m(z)\cdot\La\Phi\ =\ \sum_{i=1}^2\,{\del\over\del
x_i}P^\Phi_i\,(x_1,x_2)\ \cong\ 0\ ,\cr \I_2(\Phi)
&=\coeff1{2\pi i}\!\oint\! dz\,z^2\,\Um_m(z)\cdot\La\Phi\ =\
\coeff{\a_m}{\left( t-4 \right) \left( t+2 \right) }\ \times \cr
&\Big\{\Big[ \left({t^2}+4t-24 \right){\del\over\del x_1} + 2 \left(
3 t-8 \right) x_2{\del^2\over\del x_1\del x_2} + \left( 3 t-8 \right)
x_1{\del^2\over\del {x_1}^2}\Big]\,P^\Phi_1\,(x_1,x_2) \cr+&\Big[
\left( {t^2}-6t+12 \right){\del\over\del x_2}+ \left( t+2 \right)
x_2{\del^2\over\del {x_2}^2}- 2 \left( t-5 \right) x_1
{\del^2\over\del x_1\del x_2} \cr & \qquad\qquad\qquad \qquad\qquad \
- \left( 5 t-18 \right) {\del^2\over\del
{x_1}^2}\Big]\,P^\Phi_2\,(x_1,x_2)\Big\}\ \cong\ 0\ .\cr
}\eqn\UGaction
$$
Obviously, the operators $\I_i(\Phi)$ map between LG
polynomials, and more specifically, one has the
following system of descent equations on the space of polynomials:
$$
\eqalign{
\Phi\ &=:\
\Phi_{(0,0)} +
\sum\, P^\Phi_{i,(0,0)}\,{\del\over\del x_i}W_k \cr
\vdots\ \ &\ \ \ \ \qquad\qquad\qquad\qquad \vdots \cr
\I_1(\Phi_{(n_1,n_2)})\ &=:\
\Phi_{(n_1+1,n_2)} +
\sum\, P^\Phi_{i,(n_1+1,n_2)}\,{\del\over\del x_i}W_k \cr
\I_2(\Phi_{(n_1,n_2)})\ &=:\
\Phi_{(n_1,n_2+1)} +
\sum\, P^\Phi_{i,(n_1,n_2+1)}\,{\del\over\del x_i}W_k\ , \cr
}\eqn\descent
$$
where $\Phi_{(n_1,n_2)}\in\rx k$. Note that these expansions are in
general not unique, and this, in fact, corresponds to choosing a
gauge (that is, by writing $\I_1(\Phi_i\cdot\Phi_j)\equiv
\Gamma_{ij}^k\Phi_k$, we see that it amounts to a gauge choice for
the Gau\ss-Manin connection $\Gamma$ \flatco). Therefore the precise
image of a given $\Phi$ under the map $\I_1$ or $\I_2$ depends, in
general, on this gauge. By choosing the $P^\Phi_i$ appropriately, the
$\I_i$ can always be made to act between the $W_3$ eigen-polynomials
$\Phi^{l,m}$.

By definition, the $\I_i$ map physical operators into physical ones
and unphysical (null) operators into unphysical ones, and this allows
to recursively determine whether any given polynomial $\Phi^{l,m}$
describes a physical state or not. For $l\geq t$, it is easy to see
that one can always choose the $P^\Phi_i$ such that
$$
\I_{1,2}\ :\ \ \Phi^{l,m}\xy\ \longrightarrow\ \Phi^{l-t,m}\xy\ ,
\eqn\Imap
$$
so that all operators with $l\geq t$ can be related to operators in
the strip $l<t$. The action of the $\I_i$ on the operators with $l<t$
is, on the other hand, in general quite complicated, but can
be deduced case by case. The result from such an analysis is that all
$\Phi^{l,m}$ become physical operators after coupling to $W$-gravity,
except for
$$
\eqalign{
&\big\{\,\Phi^{(l+1)\,t-m-2,m}\xy,\ \ l=0,1,2,\dots,\
m=0,1,2,\dots(l+1)\,t-2\,\big\}\cr
&\big\{\,\Phi^{(l+1)\,t-1,m}\xy,\,\Phi^{l,(m+1)\,t-1}\xy,\ \
l,m=0,1,2,\dots\big\}\ ,\cr
}\eqn\nullop
$$
which remain null. This list corresponds to eq.\ \gx\ for one
variable.
The spectrum for $t=7$ is depicted in Fig.3, which
is the two-variable analog of Fig.1.
\ifig\figtwo{Gravitational chiral ring associated with the
topological $W_3$ minimal model $\cph24$, represented in terms of \LG
polynomials $\Phi^{l,m}\xy$. The projection on the $m$-axis gives the
$U(1)$ charge \charge. The open dots denote null fields. The leftmost
triangle of black dots represents the primary chiral ring $\rx 4$ of
the matter model, while the other black dots represent their
gravitational descendants. The descendant generators are linear
combinations of the two operators indicated.}
{\epsfxsize3.0in\epsfbox{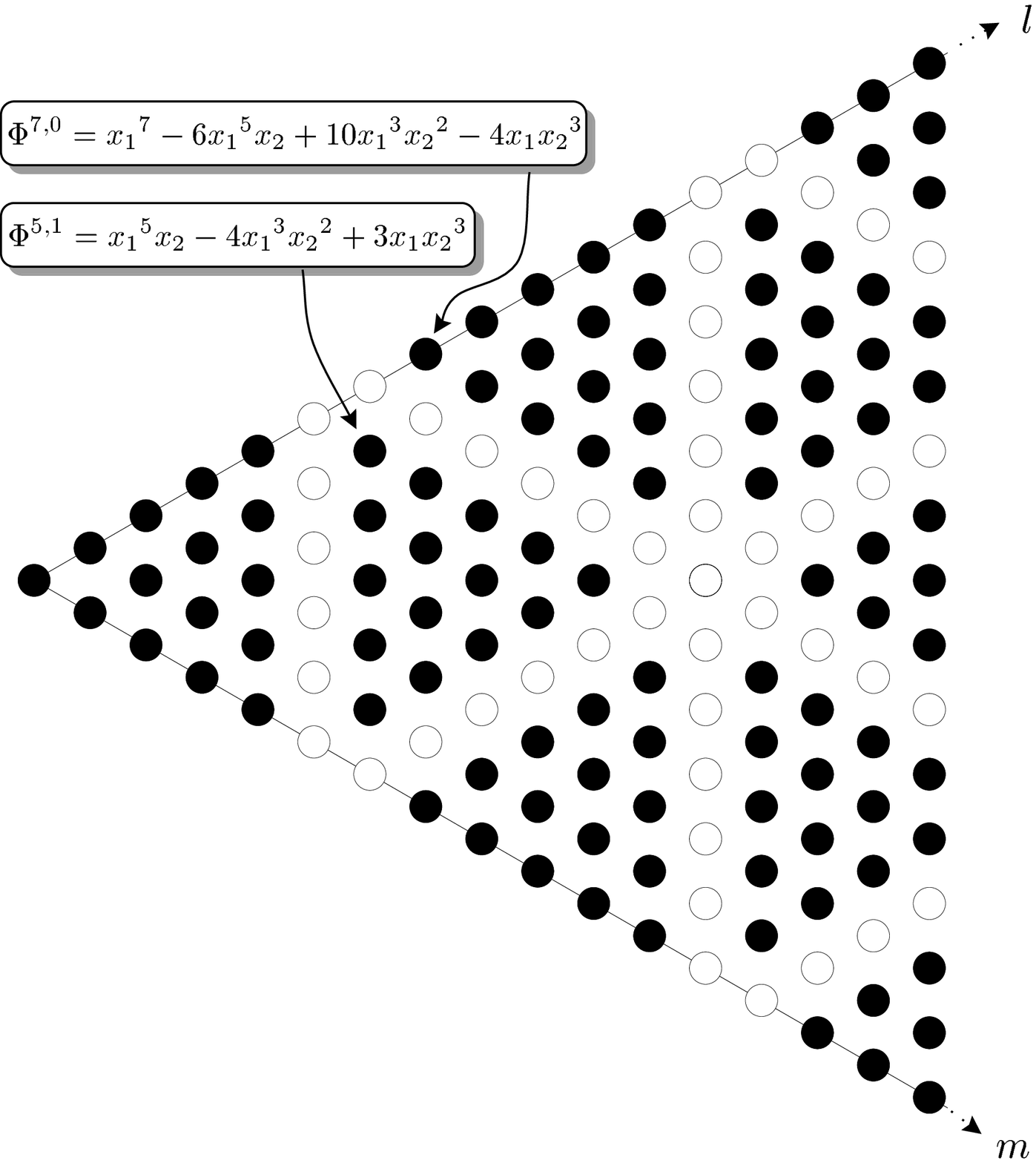}}

The interpretation of the physical states with $l+m>k\equiv t-3$ is,
of course, that they are the gravitational descendants of the matter
chiral ring $\rx k$, and it is easy to see that the $\I_i(\Phi)$ are
linear combinations of inverses of gravitational descendant
generators $\sigma_1,\sigma_2$. More specifically, consider
$$
\eqalign{
\sigma_1\ \equiv\ \sigma_1(1)\ :=\ \Phi^{t,0}\ &\equiv\ \dx1 \wk{k+1}
\cr &=\ (x_1\dx1+x_2\dx2)\,\wk k\ ,
}\eqn\sigone
$$
where in the second line we used \recVR. The expression in the
second line is the same as the expression for the dilaton in
terms of LG fields that was found by Lossev \Loss. Indeed, $\I_1$ is
identical to the recursion operator of \Loss\ that satisfies
$$
\I_1(\sigma_1(\Phi))\ =\ 2\Phi\ ,
\eqn\Ionerec
$$
which is solved by
$$
\sigma_1(\Phi)\ \cong\
\dx1\wk k\big(\int^{x_1}\!\!dx_1'\,\Phi\big)+
\dx2\wk k\big(\int^{x_2}\!\!dx_2'\,\Phi\big)\ .
\eqn\siggeneral
$$
$\I_1$ also appears in the puncture operator contact term.
Specifically, noting that for $l,m\geq0$:
$$
\I_1(\Phi^{t+l,m})\ =\ (l+m+2)\,\Phi^{l,m}\ ,
\eqn\IoneruleA
$$
one gets
$$
\int_{|z|<\epsilon}\!\!\Phi(z)\,|\,\Phi^{t+l,m}\,\rangle\ =\
\Phi(0)\,|\,\I_1(\Phi^{t+l,m})\,\rangle\ =\
(l+m+2)\,|\,(\Phi\cdot\Phi^{l,m})\,\rangle\ , \eqn\contact
$$
and this leads, essentially, to a Virasoro algebra \EHV. It would be
very interesting to find a $W$-geometric interpretation of the other
operator, $\I_2$, in relation with contact terms, but this is beyond
our present scope.

As for the other gravitational descendant generator, $\sigma_2$, we
find that using identities like \recVR\ one has for $l+m\leq k$:
$$
\I_1(\Phi^{t-l-2,m+l+1})\ = -(m+1)\,\Phi^{l,m}\ .
\eqn\IoneruleB
$$
(In terms of Fig.2, the
operators $\Phi^{t+l,m}$ and $\Phi^{t-l-2,m+l+1}$ live in the two
triangles that are pointed to by the arrows.) Therefore the
linear combinations
$$
\sigma_2(\Phi^{l,m})\ :\equiv\
(l+m+2)\,\Phi^{t-l-2,m+l+1}+(m+1)\,\Phi^{t+l,m}
\eqn\sigtwodef
$$
obey
$$
\I_1(\sigma_2(\Phi^{l,m}))\ =\ 0\ .
\eqn\siga
$$
On the other hand, one can check that
$$
\I_2(\sigma_2(\Phi^{l,m}))\ =\ v_2\,\Phi^{l,m}\ ,
\eqn\sigb
$$
with some generically non-zero coefficient $v_2$. This means that
although $\sigma_2(\Phi^{l,m})$ are null operators in ordinary
topological gravity, they become physical if one imposes the
additional $W_3$ equivariance condition, $b_2\Phi\cong 0$. Hence,
these operators can be viewed as LG representatives of the extra
gravitational descendants of $\rx k$ that are due to the $W_3$
extension of topological gravity. In particular, the $W_3$
descendant generator is
$$
\sigma_2\ \equiv\ \sigma_2(1)\ =\ \Phi^{t,0}+2 \Phi^{t-2,1}\ =\
(x_1\dx1+3x_2\dx2)\,\wk k\ .
\eqn\sigamtwo
$$
One expects that $\sigma_{1,2}$ are LG representatives of the basic
operators of topological $W_3$-gravity, and we like to find the
precise relationship by making use of an argument similar to the one
of ref.\ \EYQ. For this, note that in analogy to ordinary gravity
\doubref\topgr\topw, the non-trivial cohomology in pure topological
$W$-gravity is (supposedly completely) generated by
$$
\eqalign{
\g^0_i\ &=\ \shalf\big\{\,Q_S-\bar
Q{}_S,\,\big\{\,Q_V,\,\phi_i\,\big\}\,\big\}\cr
&=:\shalf\big\{\,Q_S,\,\La{\g_i}\big\} - c.c\ ,\ \ \ \ i=1,2.
}\eqn\gammdef
$$
(These operators are non-trivial because the Liouville fields
$\phi_i$ do not create states in the physical Hilbert space.) From
the BRST operator \Qvdef\ we have
$$
\eqalign{
\La{\g_1}\ &=\ -\shalf\del c_1-\coeff{19}{46}\del^2c_2
+\big(c_1\del\phi_1+\dots\big)\cr \La{\g_2}\ &=\ -\del
c_1+\coeff{4}{23}\del^2c_2 +\big(c_1\del\phi_2+\dots\big)\ ,\cr
}\eqn\lamgam
$$
where the parentheses indicate terms that are not important here. For
convenience, let us first introduce
$$
\tilde\sigma_1\ =\ (18-5t)\sigma_1 + (t^2-t-6)\sigma_2\ ,
\eqn\sigtilde
$$
and note that
$$
\eqalign{
\I_1(\tilde\sigma_1)\ &=\ 2(18-5t)\cr
\I_2(\tilde\sigma_1)\ &=\ 0\cr
\I_1(\sigma_2)\ &=\ 0\cr
\I_2(\sigma_2)\ &=\ 2\coeff{5t-18}{(t+2)(t-4)}\a_m\ ,\cr
}\eqn\Iobserve
$$
as well as
$$
\eqalign{
(b_1)_0\cdot\La{\g_1}\ &=\ -\shalf\cr
(b_2)_0\cdot\La{\g_1}\ &=\ -\coeff{19}{23}\cr
(b_1)_0\cdot\La{\g_2}\ &=\ -1\cr
(b_2)_0\cdot\La{\g_2}\ &=\ \coeff{8}{23}\ .\cr
}\eqn\bobserve
$$
Then, observing that the Liouville and ghost parts in
$\Gm_{tot},\Um_{tot}$ do not contribute to the constant terms that we
are looking at and that $S\g_i^0S^{-1}=\g_i^0$, we see the following
equivariance conditions to be satisfied:
$$
\eqalign{
\big\{\,(b_1 + \Gm_{tot})_{\,0},\,\a_1(8\La{\g_1}+19\La{\g_2})\,-\,
\La{\tilde\sigma_1}\,\big\}\ &=\ 0 \cr \big\{\,(b_2 +
\Um_{tot})_{\,0},\,\a_1(8\La{\g_1}+19\La{\g_2})\,-\,
\La{\tilde\sigma_1}\,\big\}\ &\equiv\ 0\ , \cr
}\eqn\firstident
$$
where $\a_1=\coeff2{23}(5t-18)$, and
$$
\eqalign{
\big\{\,(b_1 + \Gm_{tot})_{\,0},\,\a_2(2\La{\g_1}-\La{\g_2})\,-\,
\La{\sigma_2}\,\big\}\ &\equiv\ 0 \cr \big\{\,(b_2 +
\Um_{tot})_{\,0},\,\a_2(2\La{\g_1}-\La{\g_2})\,-\,
\La{\sigma_2}\,\big\}\ &=\ 0\ , \cr
}\eqn\secident
$$
where $\a_2=\coeff{18-5t}{(t+2)(t-4)}\a_m$. This means that we have,
in equivariant cohomology, the following relationship between the two
representations of the gravitational ring generators:
$$
\eqalign{
\tilde\sigma_1\xy\ &\cong\ \a_1\,( 8\g^0_1 + 19\g^0_2)\cr
\sigma_2\xy\ &\cong\ \a_2\,( 2\g^0_1 -\g^0_2) \ .\cr
}\eqn\finalident
$$
Because of the polynomial ring structure, it is quite clear that
similar equivalences should hold also for powers of the $\sigma_i$.

\chapter{Conclusions}

We investigated how topological $W$-gravity can be realized in terms
of twisted \nex2 \LG theory. We showed that by using only information
that is intrinsic to a given LG ``matter'' model, one can recover the
chiral ground ring of topological $W$-gravity, and generate pictures
like Fig.4. We considered as example the special case of $W_3$, but
it is fairly obvious that our methods should work for general $W_n$
in an analogous way (and probably for even more general LG theories
as well). \ifig\figthree{Polynomial representation of pure
$W_3$-gravity. The spectrum corresponds to the ``trivial''
superpotential $W_0=\coeff13 {x_1}^3-x_1x_2$ with vanishing \nex2
central charge. The marked dots denote physical states that were
unexpected from previous work.}{\epsfxsize3.in\epsfbox{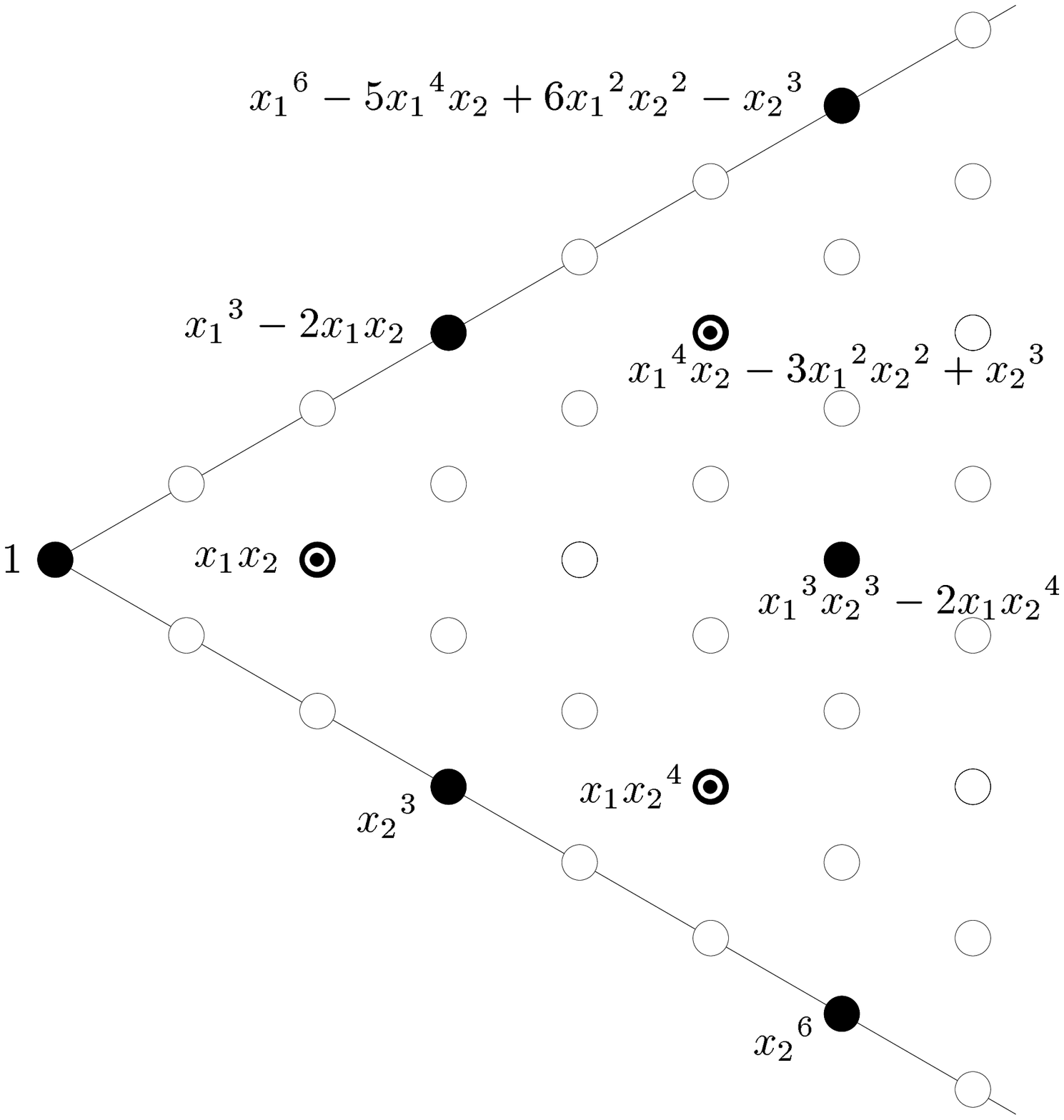}}

More specifically, we first showed how one can construct the extended
(twisted) \nex2 current algebra in terms of \LG data, which are the
LG fields and their superpotential. We then used these currents to
describe the effect of requiring equivariant cohomology on the LG
spectrum. The precise form of the gravity sector, and hence of the
BRST operator, is not important here, since it decouples. The only
important ingredient is eq.\ \intrig, which ensures that the
cohomology can be represented entirely in the LG sector. The precise
form of the chiral algebra currents then directly translates into
which LG polynomials become physical and which stay null after
coupling to topological $W$-gravity.

We find this simple algorithm quite remarkable, since the polynomial
LG spectrum obtained in this way carries implicitly a lot of
mathematical information, for example about the structure of
singular vectors of type $(1,t)$ $W_3$ minimal models, and
about representation theory of affine algebras.

Our results also indicate that the spectrum of $W$-gravity coupled to
matter contains more states than previously thought. In refs.\
\doubref\BLNWB\BMP, the $W_3$ chiral ground ring for $t=k+3$ was
found to contain the following operators:
$$
\cR_{x,\g}^{(k)}\ =\ \rx
k\xy\,\otimes\,\big\{\,(\sigma_1)^{r_1}(\sigma_2)^{r_2}\,,\ \
r_1,r_2=1,2,\dots\,\big\}\ , \eqn\oldRxg
$$
where $\sigma_i$ denote gravitational generators with $U(1)$ charges
$$
J_0(\sigma_i)\ =\ i\ .\eqn\Jyi
$$
In contrast, we find in the present paper a ring generator
$\sigma_2$ already at level one and not at level two,
$$
J_0(\sigma_i)\ =\ 1\ .\eqn\Jsi
$$
The new spectrum contains the old one, and to be
explicit, we marked in Fig.4 the new additional states.

Note that our present findings do not invalidate the results of
refs.\ \doubref\BLNWB\BMP, where one was looking for
ring generators with the simplest possible structure (that turned
out be be already quite complicated!). It would be very interesting
to understand the emergence of the additional states in terms
of the cohomology of the $W_3$-matter-gravity system.

\goodbreak \ack
We thank Erik Verlinde for discussions, and Kris Thielemans
for his Mathematica package SOPEN2defs.m.


\append {A}{Superspace conventions and the N=2 W$_{\!{\textstyle{\bf
3}}}$ algebra}

We use \nex2 superspace with coordinates $\{ z, \theta^+,
\theta^-\}$. The derivatives are defined as
$$
\eqalign{
D_\pm&\equiv D^\mp\equiv {\del\ \over\del \theta^\pm}-\half
\theta^\mp {\del\over\del z}\cr \del&\equiv {\del\over\del z}\cr
D_{\!+-}&\equiv-\shalf [D_{\!+} , D_{\!-} ]\ ,\cr
}\eqn\ssder
$$
such that
$$
\{ D_{\!+} ,D_{\!-} \} = -\del\ .
$$
{}From
$$
\coeff{1}{2\pi i}\oint dZ_1 {\tp_{12}\tm_{12}\over z_{12}}\Phi
(Z_1)=\Phi (Z_2)
$$
one deduces the \nex2 Taylor expansion
$$
\Phi (Z_1)=\sum_{n\geq 0}{z_{12}^n\over n!}\del^n\left( 1+\tp_{12}
D_{\!+}+\tm_{12} D_{\!-} + \tp_{12}\tm_{12} D_{\!+-}\right) \Phi
(Z_2),
$$
where
$$
\eqalign{\theta_{12}^\pm&\equiv\theta^\pm_1-\theta^\pm_2\cr
z_{12}&\equiv z_1-z_2+\shalf \theta^+_1\theta^-_2 + \shalf
\theta^-_1\theta^+_2.}
$$
The super energy momentum tensor $\cJ$ is an unconstrained superfield
which satisfies the OPE:
$$
\cJ(Z_1)\,\cJ(Z_2)={c\over 3} {1\over z_{12}^2}+\Bigg(
{\tp_{12}\tm_{12}\over z_{12}^2} +{\tm_{12}\over
z_{12}}D_{\!-}-{\tp_{12}\over z_{12}}D_{\!+}+ {\tp_{12}\tm_{12}\over
z_{12}}\del\Bigg)\cJ(Z_2).\eqn\TT
$$
An unconstrained superfield $\cO$ is primary with conformal dimension
$h$ and $U(1)$ charge $q$ if
$$
\cJ(Z_1)\,\cO(Z_2)=\Bigg( h {\tp_{12}\tm_{12}}{1\over z_{12}^2}
+{\tm_{12}\over z_{12}}D_{\!-}-{\tp_{12}\over z_{12}} D_{\!+}+
{\tp_{12} \tm_{12} \over z_{12}}\del+ q {1\over z_{12}}
\Bigg)\cO(Z_2).\eqn\TW
$$
The superfield $\cO$ is said to be chiral if it satisfies
$D_{\!+}\cO=0$
and consistency of eq. \TW with this requires that $h=q/2$.
Similarly, one can define an anti-chiral superfield $\cO$ which
satisfies $D_{\!-}\cO=0$ and which has $h=-q/2$.

In this paper, we mainly deal with the \nex2 $W_3$ algebra \Wchiral.
This algebra is generated by the super energy momentum tensor and one
additional unconstrained superfield $\cV$. It is primary with
conformal dimension 2 and vanishing $U(1)$ charge. The operator
product expansion of $\cV$ with itself is rather involved, but we
give it here to correct some slight misprints of the component
results given in the literature:
\ifx\answ\bigans
$$
\eqalign{
\cV(&Z_1) \,\cV(Z_2)={1\over z_{12}^4}\,\Big\{ {c\over 6} +
\tp_{12}\, \tm_{12} \, \cJ \Big\}-{1\over z_{12}^3}\,\Big\{
\tp_{12}\, D_{\!+}\cJ - \tm_{12}\, D_{\!-}\cJ - \tp_{12}\,
\,\tm_{12}\, \del\cJ \Big\}+\cr &{1\over z_{12}^2}\,\Big\{ (1
-c)^{-1}\Big( -\coeff{2 c}{3}D_{\!+-}\cJ + \cJ\cJ + \tm_{12}\,\big(
\coeff{3-2c}{3} \del D_{\!-}\cJ + \cJ D_{\!-} \cJ\big) -\cr
&\tp_{12}\,\big( \coeff{3-2c}{3} \del D_{\!+}\cJ - \cJ D_{\!+}
\cJ\big) + (4(3-2c)(6+c))^{-1}\tp_{12}\, \tm_{12}\,
\big(3(10c-24)c\del D_{\!+-}\cJ + \cr &4(36 - 9c + 8c^2)\cJ
D_{\!+-}\cJ - 12(3 + 4c)\cJ\cJ\cJ + 12(12 - 5c)c D_{\!-}\cJ
D_{\!+}\cJ+ \cr &2(18 - 15c + 2c^2 + 2c^3)\del^2 \cJ\big)\Big)+
\alpha\Big( \cV + \shalf \tm_{12}\, D_{\!-}\cV + \shalf \tp_{12}\,
D_{\!+}\cV - \cr &\tp_{12}\, \tm_{12}\,3(4(12-5c))^{-1}\big( 28
\cJ\cV-2 (8-c)D_{\!+-} \cV\big) \Big) \Big\}+\cr &{1\over
z_{12}}\,\Big\{ (1-c)^{-1}\del \cJ\cJ-c(3(1-c))^{-1}\del
D_{\!+-}\cJ+\cr &\tp_{12}\, (2(1-c)(3-2c)(6+c))^{-1}\big( -c(9 - 3 c
+ c^2)\del^2D_{\!+}\cJ + 6(3 + 4c)\cJ\cJ D_{\!+}\cJ+\cr &18(1 - c)c
D_{\!+-}\cJ D_{\!+}\cJ + 2(18 - 24 c - c^2)\del D_{\!+}\cJ\cJ - 3 (1
- c)(6 - c)\del\cJ D_{\!+}\cJ\big) +\cr &\tm_{12}\,
(2(1-c)(3-2c)(6+c))^{-1}\big(\shalf c(9 + 3 c + 2 c^2)\del^2
D_{\!-}\cJ - 6(3 + 4c)\cJ\cJ D_{\!-}\cJ-\cr &18(1 - c)c D_{\!-}\cJ
D_{\!+-} \cJ + 2(18 - 24 c - c^2)\del D_{\!-}\cJ\cJ - 3 (1 - c)(6 -
c)\del\cJ D_{\!-}\cJ\big) +\cr &\tp_{12}\, \tm_{12}\,
(2(1-c)(3-2c)(6+c))^{-1}\big( 4(12 - 5c)c\del D_{\!-}\cJ
D_{\!+}\cJ+\cr &8(9 - 3 c + c^2)\del D_{\!+-}\cJ\cJ- 4(12-5c)c \del
D_{\!+}\cJ D_{\!-}\cJ + 4c(3 + 4c) \del\cJ D_{\!+-}\cJ - \cr &12(3 +
4c) \del\cJ\cJ\cJ - \coeff{1}{3} (18 +3 c + 2c^2 -
2c^3)\del^3\cJ\big) + \alpha \Big(\shalf \del\cV +\cr &\tp_{12}\,
(2(12-5c)(3+c))^{-1}\big( 3(c-1)(6-c)\del D_{\!+}\cV + 6(15 - c) \cJ
D_{\!+}\cV-\cr &54(1 - c)D_{\!+}\cJ\cV\big)+ \tm_{12}\,
(2(12-5c)(3+c))^{-1}\big( 3(c-1)(6-c)\del D_{\!-}\cV - \cr &6(15 - c)
\cJ D_{\!-}\cV+ 54(1 - c)D_{\!-}\cJ\cV\big)+ \tp_{12}\, \tm_{12}\,
(2(12-5c)(3+c))^{-1}\times\cr &\big(2(15 - c)c \del D_{\!+-}\cV -
18(6 + c)\cJ\del\cV + 6(12 - 5c)D_{\!-}\cJ D_{\!+}\cV + \cr &6(12 -
5c)D_{\!+}\cJ D_{\!-}\cV - 12(3 + 4c)\del\cJ\cV\big)\Big) \Big\}\cr
}\eqn\WW
$$
\else
$$
\eqalign{
\cV(&Z_1) \,\cV(Z_2)={1\over z_{12}^4}\,\Big\{ {c\over 6} +
\tp_{12}\, \tm_{12} \, \cJ \Big\}-{1\over z_{12}^3}\,\Big\{
\tp_{12}\, D_{\!+}\cJ - \tm_{12}\, D_{\!-}\cJ - \tp_{12}\,
\,\tm_{12}\, \del\cJ \Big\}+\cr &{1\over z_{12}^2}\,\Big\{ (1
-c)^{-1}\Big( -\coeff{2 c}{3}D_{\!+-}\cJ + \cJ\cJ + \tm_{12}\,\big(
\coeff{3-2c}{3} \del D_{\!-}\cJ + \cJ D_{\!-} \cJ\big) -\cr
&\tp_{12}\,\big( \coeff{3-2c}{3} \del D_{\!+}\cJ - \cJ D_{\!+}
\cJ\big) + (4(3-2c)(6+c))^{-1}\tp_{12}\, \tm_{12}\,
\big(3(10c-24)c\del D_{\!+-}\cJ + \cr &4(36 - 9c + 8c^2)\cJ
D_{\!+-}\cJ - 12(3 + 4c)\cJ\cJ\cJ + 12(12 - 5c)c D_{\!-}\cJ
D_{\!+}\cJ+ \cr &2(18 - 15c + 2c^2 + 2c^3)\del^2 \cJ\big)\Big)+
\alpha\Big( \cV + \shalf \tm_{12}\, D_{\!-}\cV + \shalf \tp_{12}\,
D_{\!+}\cV - \cr &\tp_{12}\, \tm_{12}\,3(4(12-5c))^{-1}\big( 28
\cJ\cV-2 (8-c)D_{\!+-} \cV\big) \Big) \Big\}+\cr &{1\over
z_{12}}\,\Big\{ (1-c)^{-1}\del \cJ\cJ-c(3(1-c))^{-1}\del
D_{\!+-}\cJ+\cr &\tp_{12}\, (2(1-c)(3-2c)(6+c))^{-1}\big( -c(9 - 3 c
+ c^2)\del^2D_{\!+}\cJ + 6(3 + 4c)\cJ\cJ D_{\!+}\cJ+\cr &18(1 - c)c
D_{\!+-}\cJ D_{\!+}\cJ + 2(18 - 24 c - c^2)\del D_{\!+}\cJ\cJ - 3 (1
- c)(6 - c)\del\cJ D_{\!+}\cJ\big) +\cr &\tm_{12}\,
(2(1-c)(3-2c)(6+c))^{-1}\big(\shalf c(9 + 3 c + 2 c^2)\del^2
D_{\!-}\cJ - 6(3 + 4c)\cJ\cJ D_{\!-}\cJ-\cr &18(1 - c)c D_{\!-}\cJ
D_{\!+-} \cJ + 2(18 - 24 c - c^2)\del D_{\!-}\cJ\cJ - 3 (1 - c)(6 -
c)\del\cJ D_{\!-}\cJ\big) +\cr &\tp_{12}\, \tm_{12}\,
(2(1-c)(3-2c)(6+c))^{-1}\big( 4(12 - 5c)c\del D_{\!-}\cJ
D_{\!+}\cJ+\cr &8(9 - 3 c + c^2)\del D_{\!+-}\cJ\cJ- 4(12-5c)c \del
D_{\!+}\cJ D_{\!-}\cJ + 4c(3 + 4c) \del\cJ D_{\!+-}\cJ - \cr &12(3 +
4c) \del\cJ\cJ\cJ - \coeff{1}{3} (18 +3 c + 2c^2 -
2c^3)\del^3\cJ\big) + \alpha \Big(\shalf \del\cV +\cr
}$$ \goodbreak $$\eqalign{
&\tp_{12}\,
(2(12-5c)(3+c))^{-1}\big( 3(c-1)(6-c)\del D_{\!+}\cV + 6(15 - c) \cJ
D_{\!+}\cV-\cr &54(1 - c)D_{\!+}\cJ\cV\big)+ \tm_{12}\,
(2(12-5c)(3+c))^{-1}\big( 3(c-1)(6-c)\del D_{\!-}\cV - \cr &6(15 - c)
\cJ D_{\!-}\cV+ 54(1 - c)D_{\!-}\cJ\cV\big)+ \tp_{12}\, \tm_{12}\,
(2(12-5c)(3+c))^{-1}\times\cr &\big(2(15 - c)c \del D_{\!+-}\cV -
18(6 + c)\cJ\del\cV + 6(12 - 5c)D_{\!-}\cJ D_{\!+}\cV + \cr &6(12 -
5c)D_{\!+}\cJ D_{\!-}\cV - 12(3 + 4c)\del\cJ\cV\big)\Big) \Big\}\cr
}\eqn\WW
$$
\fi
where
$$
\alpha={\sqrt{2}(3 + c)(12 - 5c)\over
  \sqrt{3(3 - 2c)(-15 + c)(-1 + c)(6 + c)}}\ .
$$
\refout
\end